\documentclass[letter,12pt]{article}

\DeclareMathAlphabet\mathbfcal{OMS}{cmsy}{b}{n}
\usepackage[english]{babel}

\usepackage[letterpaper,top=2.54cm,bottom=2.54cm,left=2.54cm,right=2.54cm]{geometry}

\usepackage{amsmath}
\usepackage{mathrsfs}
\usepackage{graphicx}
\usepackage{subfig}
\usepackage{setspace}
\usepackage[rflt]{floatflt}
\usepackage{tikz}
\usepackage{amssymb}
\usepackage[titletoc]{appendix}

\newcommand{\We}{W\!e}
\newcommand{\bphi}{\mbox{\boldmath$\phi$}}

\newcommand{\bpsi}{\mbox{\boldmath$\psi$}}
\newcommand{\bxi}{\mbox{\boldmath$\xi$}}
\newcommand{\bkappa}{\mbox{\boldmath$\kappa$}}

\newcommand{\DefinedAs}[0]{\mathrel{\mathop:}=}
\DeclareMathOperator*{\trace}{trace}

\begin{document}
\title{\bf Amplification of localized body forces in channel flows of viscoelastic fluids}
\author{Gokul Hariharan,$^\text{a}$ Mihailo R. Jovanovi\'c$,^{\text{b}}$ and Satish Kumar$^{\text{a}}$}
\maketitle
\begin{center}
$^a$Department of Chemical Engineering and Materials Science, University of Minnesota,\newline Minneapolis, MN 55455, USA
\\[0.25cm]
$^b$Ming Hsieh Department of Electrical Engineering, University of Southern California,\newline Los Angeles, CA 90089, USA 
\end{center}
\newcommand{\D}[2]{\frac{\partial #1}{\partial #2}} 
\newcommand{\DD}[2]{\frac{\partial^2 #1}{\partial #2^2}}
\newcommand{\BB}[1]{\boldsymbol #1} 
\newcommand{\hh}[1]{\mathbf{\bar{\text{$#1$}}}}
\newcommand{\MM}[1]{\mathcal{#1}}
\newcommand{\MMbf}[1]{\mathbfcal{#1}}
\begin{abstract}
The study of nonmodal amplification of distributed body forces in channel flows of viscoelastic fluids has provided useful insights into the mechanisms that may govern the initial stages of transition to elastic turbulence. However, distributed body forces are not easy to implement in experiments and so there is a need to examine amplification of localized body forces. In this work, we use the linearized governing equations to examine such amplification in Poiseuille flow of FENE-CR fluids. We first identify the wall-normal location at which the impulsive excitations experience the largest amplification and then analyze the kinetic energy of the fluctuations and the resulting flow structures. For both a Newtonian fluid at high Reynolds numbers and a viscoelastic fluid at low Reynolds numbers, the largest amplification occurs for disturbances that are located near the channel wall. Our analysis of the energy evolution shows that a localized body force in the spanwise direction has the largest impact and that the streamwise velocity component is most affected. For viscoelastic fluids we observe the development of vortical structures away from the source of impulsive excitation. This feature is less prominent in Newtonian fluids and it may provide a mechanism for triggering the initial stages of transition to elastic turbulence.
\end{abstract}

	\vspace*{-2ex}
\section{Introduction}\label{sec:intro}

Seminal work by Groisman and Steinberg has demonstrated that dilute polymer solutions can produce a turbulent-like flow state at low Reynolds numbers~\cite{groisman2000elastic}. Such a flow state is called elastic turbulence and it has high potential for enhancing mixing~\cite{PhysRevE.69.066305} and heat transport~\cite{Copeland2017} in microfluidic flows. It can also be used to produce nonlinear effects to build microscale control devices including nonlinear flow resistors and flow memory devices such as flip-flops analogous to those in electric circuits~\cite{groisman2003microfluidic}. However, elastic turbulence is not desired in certain industrial applications, e.g., those involving polymer processing and coating flows~\cite{larson1992instabilities,MorozovSubcritWeakly}.

Elastic turbulence observed in the experiments of Groisman and Steinberg~\cite{groisman2000elastic} is thought to have originated from linear instability of curved streamlines to small-amplitude perturbations. Even though analysis of the linearized governing equations predicts stability of inertialess channel flows with straight streamlines~\cite{larson1992instabilities}, recent experiments suggest that elastic turbulence can also occur in such flows~\cite{pan2013nonlinear,qin2016elastic,Bonn2011}. This is a puzzling observation with both fundamental and technological ramifications.  For example, polymer processing operations often involve flows through straight channels and instabilities at low Reynolds numbers are detrimental to the quality of the final products~\cite{larson1992instabilities}. Furthermore, as indicated above, triggering elastic turbulence also finds positive applications in microfluidic devices. 

The absence of linear modal instability does not preclude the possibility that the early stages of transition to elastic turbulence can be understood via analysis of the linearized equations. Nonmodal analysis considers the possibility that flow fluctuations that decay asymptotically can grow transiently and that exogenous disturbances can be significantly amplified by the underlying dynamics~\cite{jovbamJFM05,schmid2007nonmodal,Trefethen}. Disturbances that experience linear nonmodal amplification can generate finite-amplitude perturbations that may trigger nonlinear flow states and induce transition to elastic turbulence. 

In refs. \cite{hodjovkumJFM08} and \cite{hodjovkumJFM09}, it was demonstrated that distributed body forces can experience significant nonmodal amplification in Couette and Poiseuille flows at low Reynolds numbers when viscoelastic effects are strong. This work showed that streamwise-constant flow structures in Oldroyd-B fluids become increasingly prominent with an increase in viscoelastic effects. This inspired Jovanovi\'c and Kumar to closely examine dynamics of streamwise-constant fluctuations in weakly inertial channel flows of viscoelastic fluids~\cite{jovkumPOF10,jovkumJNNFM11}. Their work showed that nonmodal amplification arises from a coupling between the base-state stresses and flow fluctuations, demonstrated the existence of a viscoelastic analogue of the well-known inertial lift-up mechanism, and established conceptual and mathematical similarities between nonmodal amplification in viscoelastic channel flows at low Reynolds numbers and in Newtonian channel flows at high Reynolds numbers. 

The joint influence of inertia and elasticity on the evolution of streamwise-elongated fluctuations in Couette flow of Oldroyd-B fluids was studied in ref. \cite{pagzakJFM2014}. The response of the linearized equation for the wall-normal vorticity in the presence of a decaying streamwise vortex was computed and different regimes were identified based on the relation between the solvent diffusion and polymer relaxation times.
The influence of finite extensibility of polymer molecules on the worst-case amplification of deterministic distributed body forces has also been examined using the FENE-CR model~\cite{liejovkumJFM13}. This work demonstrated that even in flows with infinitely large Weissenberg numbers, the finite extensibility of polymer molecules limits the largest achievable amplification. In related work, the viscoelastic equivalent of the well-known Orr mechanism was studied for both the Oldroyd-B and FENE-P models~\cite{pagzakJFM2015}.
	
The aforementioned work has provided important insights into the linearized dynamics of channel flows of viscoelastic fluids in the presence of distributed body forces. To achieve a direct correspondence between theory and experiment in the linearized setting, one would have to induce distributed body forces without significantly altering the mean flow, which is extremely challenging. Even if a distributed body force can be generated in an experiment and used as a starting point for a linearized analysis, it is still difficult to systematically segregate the different stages that lead to elastic turbulence by the introduction of \mbox{such a force.}

In contrast, localized body forces can be readily approximated in experiments and direct numerical simulations. Furthermore, flow transition arising from the introduction of a localized body force can be dissected to demonstrate the different stages of transition to nonlinear states. For this reason, localized body forces have been applied in many experimental studies of transition in Newtonian fluids at high Reynolds numbers~\cite{carlson1982flow,klingmann1992transition,Lemoult2014}. Moreover, in theoretical and computational studies, a localized body force can be approximated by a spatio-temporal impulsive excitation. As demonstrated in refs. \cite{mj-phd04} and \cite{jovbamACC01} and further expanded on in ref.~\cite{schmid2007nonmodal}, such an analysis exemplifies dominance of
streamwise-elongated structures inside the resulting wave packet in the early stages of disturbance amplification. Recently, localized body forces were used in viscoelastic channel flows to study drag-reduction at high Reynolds numbers~\cite{agarwal2014linear}.

Any spatially varying and temporally distributed body force can be expressed as a summation of impulses of different magnitudes, spatial positions, and temporal occurrences~\cite{huerre2000open}. The impulse response therefore contains useful information for characterizing responses of linear systems to exogenous  excitation sources. As noted above, previous work on Newtonian fluids has shown that by examining the influence of spatio-temporal impulsive forcing on the linearized dynamics, several features of the early stages of transition to turbulence in high-Reynolds-number flows of Newtonian fluids can be captured~\cite{mj-phd04,jovbamACC01,schmid2007nonmodal}.

In this paper, we systematically analyze the response of the linearized dynamics of a viscoelastic fluid in Poiseuille flow to a localized body force. We first identify the wall-normal location at which the impulse has the largest impact on the flow. We then analyze the evolution of the energy of velocity fluctuations arising from the point force applied at the optimal location and demonstrate that the amplification increases with an increase in polymer concentration and with an increase in the polymer relaxation time. Finally, we analyze flow structures that result from the impulse and discuss their potential role in the early stages of transition to elastic turbulence. 

The remainder of our presentation is organized as follows. In \S~\ref{sec:prob_for}, we describe the modeling and numerical methods employed in this work. In \S~\ref{sec:y0loc}, we present results pertaining to the identification of the optimal wall-normal location of the impulsive forcing. In \S~\ref{sec:Energy_Amplification}, we analyze the kinetic energy of flow fluctuations that arise from the application of a point force at the identified optimal wall-normal location. We discuss the resulting flow structures in~\S~\ref{sec: Flow Structures},  summarize our findings in \S~\ref{sec:conclusions}, and relegate background technical material to the appendices.

\vspace*{-2ex}
\section{Problem formulation}\label{sec:prob_for}

In this section, we present the governing equations in their evolution form, the numerical methods we use, and the way we characterize the kinetic energy of velocity fluctuations. We use the finitely extensible nonlinear elastic Chilcott-Rallison (FENE-CR) constitutive equation \cite{chilcott1988creeping} as it accounts for the finite extensibility of polymer molecules and exhibits a constant shear viscosity. Results obtained using the FENE-CR model thus allows us to isolate the influence of fluid elasticity.
	
		\vspace*{-2ex}
	\subsection{Evolution form of governing equations}\label{geef}
	
\setlength{\unitlength}{0.8cm} 
\begin{figure}
\begin{picture}(9,7)(-4,-1)
\linethickness{0.2mm} 
\multiput(2,0)(4,0){1}%
{\line(1,0){6}}
\multiput(4,0)(4,0){1}%
{\line(0,1){2.5}}
\multiput(2,2.5)(4,0){1}%
{\line(1,0){6}}
\multiput(1.8,1.25)(4,0){1}%
{\vector(0,1){0.9}}
\multiput(1.8,1.25)(4,0){1}%
{\vector(1,0){0.9}}
\multiput(1.8,1.25)(4,0){1}%
{\vector(-2,-3){0.4}}
\put(1.5,2.25){$y$}
\put(2.6,1.4){$x$}
\put(1,0.7){$z$}
\qbezier(4,0)(8.5,1.25)(4,2.5)
 \put(4,0.2){\vector(1,0){0.7}}
 \put(4,0.4){\vector(1,0){1.15}}
 \put(4,0.6){\vector(1,0){1.6}}
 \put(4,0.8){\vector(1,0){2}}
 \put(4,1){\vector(1,0){2.2}}
 \put(4,1.2){\vector(1,0){2.3}}
 \put(4,1.4){\vector(1,0){2.2}}
 \put(4,1.6){\vector(1,0){2.1}}
 \put(4,1.8){\vector(1,0){1.8}}
 \put(4,2){\vector(1,0){1.4}}
 \put(4,2.2){\vector(1,0){1}}
 \put(4,2.4){\vector(1,0){0.4}}
 \qbezier(6,3)(10.5,4.25)(6,5.5)
 \put(6,3.2){\vector(1,0){0.7}}
 \put(6,3.4){\vector(1,0){1.15}}
 \put(6,3.6){\vector(1,0){1.6}}
 \put(6,3.8){\vector(1,0){2}}
 \put(6,4){\vector(1,0){2.2}}
 \put(6,4.2){\vector(1,0){2.3}}
 \put(6,4.4){\vector(1,0){2.2}}
 \put(6,4.6){\vector(1,0){2.1}}
 \put(6,4.8){\vector(1,0){1.8}}
 \put(6,5){\vector(1,0){1.4}}
 \put(6,5.2){\vector(1,0){1}}
 \put(6,5.4){\vector(1,0){0.4}}
\put(7.5,0){\vector(0,1){2.5}}
\put(7.5,2.5){\vector(0,-1){2.5}}
\put(7.8,1.25){$2h$}
\put(4,2.5){\line(2,3){2}}
\put(4,0){\line(2,3){2}}
\put(6,3){\line(0,1){2.5}}
\put(6,3){\line(0,1){2.5}}
\put(2,2.5){\line(2,3){2}}
\put(8,2.5){\line(2,3){2}}
\put(4,5.5){\line(1,0){6}}
\put(8,0){\line(2,3){2}}
\put(2,0){\line(2,3){2}}
\put(4,3){\line(1,0){6}}
\end{picture}
\caption{\label{fig:1} Flow geometry and the steady-state parabolic velocity profile for Poiseuille flow.}
\end{figure}
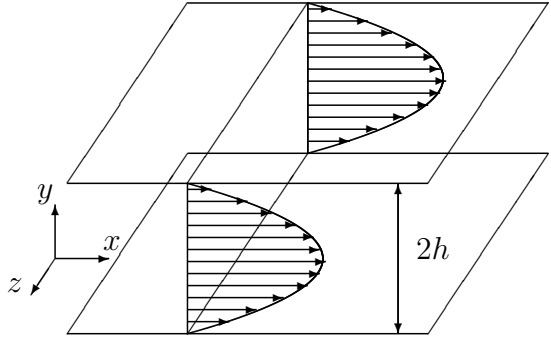

We consider a dilute polymer solution of density $\rho$ and relaxation time $\lambda$ in a channel flow whose geometry is shown in Figure \ref{fig:1}.  Length is scaled with the half-channel width $h$, velocity with the maximum velocity in the channel $U_0$, and time with $h/U_0$. Pressure is scaled with $\eta_T U_0/h$, where $\eta_T=\eta_p+\eta_s$ is the total shear viscosity with $\eta_p$ and $\eta_s$ denoting the polymer and solvent contributions to $\eta_T$. Polymer stresses are scaled with $\eta_p U_0/h$. 

This scaling leads to three non-dimensional groups: the viscosity ratio, $\beta = \eta_s/(\eta_p+\eta_s)$, the Weissenberg number, $\We = \lambda U_0/h$, and the Reynolds number, $Re = h\rho U_0/\eta_T$. The viscosity ratio provides a measure of the solvent contribution to the shear viscosity, the Weissenberg number gives the ratio of the relaxation time of the polymer to the characteristic flow time, $h/U_0$, and the Reynolds number is the ratio of inertial forces to viscous forces. In addition, the elasticity number, $\mu = \We/Re\text{ }=\lambda/(h^2\rho/\eta_T)$, determines the ratio between the fluid relaxation time and characteristic vorticity diffusion time.

The dimensionless momentum and continuity equations are
	\begin{subequations}
	 \begin{align}
	Re (\partial_t\BB{V} + \BB{V}\cdot\BB{\nabla} \BB{V}) \,&=\, -\BB{\nabla} P + \beta\nabla^2 \BB{V}+(1-\beta)\BB{\nabla} \cdot \BB{T},
	\label{eq:1a}
	\\[0.1cm]
	\BB{\nabla}\cdot\BB{V} \,&=\, 0,
	\label{eq:1b}
	\end{align}
	\label{eq:1}
\end{subequations}
where $\partial_t$ denotes a partial derivative with respect to time $t$, $\BB{V}$ is the velocity vector, $P$ is the pressure, and $\BB{T}$ is the polymer contribution to the stress tensor. 

\begin{subequations}


The conformation tensor is the mean of the dyadic product of the end-to-end vector of the finitely extensible dumbbell that is the basis of the FENE-CR model. The polymer stress tensor $\BB{T}$ is related to the conformation tensor $\BB{R}$ by
	\begin{align}
	\partial_t\BB{R} + \BB{V}\cdot \nabla \BB{R} - \BB{R}\cdot \nabla \BB{V} -(\BB{R}\cdot \nabla \BB{V})^T
\,&=\, -\BB{T},
	\label{eq:1c}
	\\
	 \frac{f}{\We}(\BB{R}-\BB{I})\,&=\,\BB{T} ,
	\label{eq:1d}
	\end{align}
where $\BB{I}$ is the identity tensor and $f$ quantifies the nonlinear spring interaction,
\begin{equation}
f = \frac{L^2-3}{L^2 - \text{trace}(\BB{R})}.
\end{equation}
We note that $\BB{R}$ and $L^2$ are scaled with $kT/c$, and $k$, $T$, and $c$ are the Boltzmann constant, absolute temperature, and spring constant of the dumbbells, respectively. As $L\rightarrow \infty$, the FENE-CR model simplifies to the Oldroyd-B model. Furthermore, system \eqref{eq:1} reduces to the Navier-Stokes equations as $\beta\rightarrow 1$.    
\end{subequations}

The steady-state solution of system \eqref{eq:1} for plane Poiseuille flow is
\begin{subequations}\label{eq:4_basestate}
\begin{align}
\bar{\BB{V}} \, &= \, \big[
\begin{array}{ccc}
\bar{U}(y) & 0 & 0
\end{array}
\big]^T,
	\\[0.1cm]
\bar{\BB{R}} \, &= 
\left[
\begin{array}{ccc}
1 + 2\text{ } (\We\text{ } \bar{U}'(y)/\bar{f})^2 & \We \text{ }\bar{U}'(y)/\bar{f} & 0\\
\We\text{ }\bar{U}'(y)/\bar{f} & 1 & 0\\
0 & 0 & 1
\end{array}
\right],
\end{align}
where 
\begin{equation}\label{eq:5}
\bar{U}(y) = 1-y^2,\quad \bar{f} = \frac{1}{2}\left(1+\sqrt{1+8\left(\frac{\We\text{ }\bar{U}'(y)}{\bar{L}}\right)^2}\right),\quad \bar{L}^2 = L^2-3.
\end{equation}
\end{subequations}
The steady-state velocity has the same parabolic profile as a Newtonian fluid because of the absence of shear-thinning effects in the FENE-CR constitutive equation. There is, however, a first normal stress difference in the FENE-CR fluid.  

The linearized equations that govern the evolution of fluctuations about the steady-state~\eqref{eq:4_basestate} are given by
\begin{subequations}
	\begin{align}
	Re\;\partial_t\BB{v}\,&=\,-\BB{\nabla}p + (1-\beta)\BB{\nabla}\cdot \BB{\tau} + \beta\nabla^2\BB{v} -Re(\bar{\BB{V}}\cdot \BB{\nabla}\BB{v} + \BB{v}\cdot\BB{\nabla}\bar{\BB{V}}) + \BB{d},
	\label{eq:6a}
	\\
	\BB{\nabla}\cdot\BB{v} \,&=\, 0,
	\label{eq:6b}
	\\
	\partial_t\BB{r}\,&=\, \BB{r}\cdot\BB{\nabla}\bar{\BB{V}} + \bar{\BB{R}}\cdot\BB{\nabla}\BB{v} + (\BB{r}\cdot\BB{\nabla}\bar{\BB{V}})^T + (\bar{\BB{R}}\cdot\BB{\nabla}\BB{v})^T-\BB{v}\cdot \BB{\nabla}\bar{\BB{R}}-\bar{\BB{V}}\cdot \BB{\nabla}\BB{r}-\BB{\tau},
	\label{eq:6c}
	\\
	\BB{\tau}\,&=\, \frac{\bar{f}}{\We}\left(\BB{r}+\frac{\bar{f}(\bar{\BB{R}}-\BB{I})}{\bar{L}^2}\text{trace}(\BB{r})\right).
	\label{eq:6d}
	\end{align}
	\label{eq:6_lnsNV}
\end{subequations}
Here, $\BB{v}$, $p$, $\BB{r}$, and $\BB{\tau}$ denote velocity, pressure, conformation tensor, and stress tensor fluctuations about their respective base profiles, $\bar{\BB{V}}$, $\bar{P}$, $\bar{\BB{R}}$, and $\bar{\BB{T}}$. We denote the components of the velocity fluctuation vector by $\BB{v} = [ \, u ~\, v ~\, w ~\, ]^T$, where $u$, $v$, and $w$ represent the streamwise ($x$), wall-normal ($y$), and spanwise ($z$) velocities, respectively. 

	The body forcing $\BB{d}$ is used to excite flow fluctuations. In this work, we use an impulsive body force,
\begin{equation}\label{eq:13}
\BB{d} (x,y,z,t) \; = \; \delta(x,y,z,t) \, \BB{e}_i,
\end{equation}
where $\BB{e}_i$ is a unit vector in the $i$th coordinate direction and $\delta(x,y,z,t)$ is the Dirac delta function in space and time. 
 
System \eqref{eq:6_lnsNV}  can be simplified by eliminating pressure  and expressing the velocity fluctuations in terms of wall-normal velocity $v$ and vorticity $\eta \DefinedAs \partial_zu-\partial_xw$. This is done by taking the divergence of \eqref{eq:6a} to get an explicit expression for $p$. Substituting this expression for $p$ into \eqref{eq:6a} yields the equation for the wall-normal velocity and the equation for $\eta$ is determined by the $y$-component of the curl of \eqref{eq:6a}. Finally, the stress tensor can be eliminated in favor of the conformation tensor using relations \eqref{eq:6c} and \eqref{eq:6d}. 

After the above algebraic manipulations, and after taking a Fourier transform in the $x$- and $z$-directions, we obtain the following evolution form for the linearized equations,
	\begin{equation}\label{eq:11}
	\begin{split}
	\partial_t \bpsi (\bkappa,y,t) & \, = \,
	\left[ \, \mathbf{A}(\bkappa) \, \bpsi(\bkappa, \, \cdot \, ,t) \, \right] (y) 
	\, + \,
	\left[ \, \mathbf{B}(\bkappa) \, \BB{d} (\bkappa, \, \cdot \, ,t) \, \right] (y),
	\\[0.1cm]
	\bphi (\bkappa,y,t) & \, = \, 
	\left[ \, \mathbf{C}(\bkappa) \, \bpsi (\bkappa, \, \cdot \, ,t) \, \right] (y),
	\end{split}
	\end{equation}
where $\bpsi = [\, \BB{r}^T ~\, v ~\, \eta \, ]^T$ is the state with $\BB{r}$ denoting the vector of the six fluctuating components of the (symmetric) conformation tensor. The linear integro-differential operators in the wall-normal direction $\mathbf{A}$, $\mathbf{B}$, and $\mathbf{C}$ are defined in the Appendix~\ref{appA} and they map the input $\BB{d}$ (i.e., the imposed forcing) to the output $\bphi = [\, u ~\, v ~\, w \, ]^T$ (i.e., the vector of velocity fluctuations) through the evolution model. We define $\bkappa = (k_x,k_z)$, where $k_x$ and $k_z$ represent the wavenumbers in the $x$- and $z$-directions. The no-slip and no-penetration boundary conditions are applied to the wall-normal velocity and vorticity components in~\eqref{eq:11},
\begin{equation}\label{eq:12a}
v(\bkappa,y=\pm 1,t)= \partial_y v(\bkappa, y=\pm 1,t) = \eta(\bkappa,y=\pm 1,t) = 0.
\end{equation}

	\vspace*{-2ex}
\subsection{Numerical method}\label{numer_meth}

Evolution model~\eqref{eq:11} represents a system of integro-differential equations in $y$ and $t$, parametrized by the wavevector $\bkappa = (k_x,k_z)$. The wall-normal direction is discretized using a Chebyshev pseudospectral technique with $N$ collocation points to reduce \eqref{eq:11} with boundary conditions \eqref{eq:12a} to a system of ordinary differential equations (ODEs) in time. All calculations are carried out using the Matlab Differentiation Matrix Suite of Weidmann and Reddy \cite{weideman2000matlab}. 

Since the wall-normal direction is discretized on a finite grid of Chebyshev collocation points, we employ the following approximation for the Dirac delta function in $y$,
\begin{equation}\label{eq:14}
\delta_0(y) \, \approx \, \frac{1}{2\sqrt{\pi\epsilon}} \, \mathrm{e}^{- \frac{(y \, - \, y_0)^2}{4 \epsilon}},\quad \epsilon>0,
\end{equation}
where $y_0$ denotes the location of the impulse in the wall-normal direction and $\epsilon$ is a small parameter. In this work, we set $\epsilon = 1/2000$. We found that this value is sufficiently small to represent an impulse as the results do not change significantly by further reducing $\epsilon$. We discuss the choice of $y_0$ in \S~\ref{sec:y0loc}. 

The forcing term in the evolution model~\eqref{eq:11} is then given by


	\begin{equation}\label{F}
	\left[ \, \mathbf{B}(\bkappa) \,\BB{d} (\bkappa,\cdot,t) \, \right](y)	\; = \;
	\BB{F}_i (\bkappa,y) \, \delta (t),
	\end{equation}
where 
	\begin{equation}\label{Fi}
	\BB{F}_i (\bkappa,y)
	\; = \; 
	\left[ \, \mathbf{B}(\bkappa) \, {\delta_0} (\cdot) \BB{e}_i \, \right] (y)
	\end{equation}	
and ${\BB{e}}_i$ is a unit vector in the $i$th coordinate direction with $i = x$, $y$, or $z$. The resulting finite-dimensional approximation to \eqref{eq:11} is then given by
\begin{equation}
	\begin{split}
	\dot{\psi}(\bkappa,t) 
	& \, = \,
	A(\bkappa) \, \psi(\bkappa,t) 
	\; + \; 
	{F_i}(\bkappa) \, \delta (t),
	\\[0.1cm]
	\phi(\bkappa,t) 
	& \, = \,
	C(\bkappa) \, \psi(\bkappa,t),
	\end{split}
	\label{eq:12}
\end{equation}  
where $\psi(\bkappa,t)$ and $\phi(\bkappa,t)$ are complex-valued vectors with $8N$ and $3N$ entries, respectively, $A(\bkappa)$ and $C(\bkappa)$ are the finite-dimensional approximations of the corresponding operators in \eqref{eq:11}, and $F_i(\bkappa)$ is the discrete approximation to $\BB{F}_i (\bkappa,y)$ in \eqref{Fi}.    

The solution of \eqref{eq:12} with zero initial conditions arising from the impulsive excitation in the $i$th coordinate direction is given by~\cite{hespanha2009linear},
\begin{equation}\label{16b}
	\phi_i(\bkappa,t) \,=\, C (\bkappa) \int_0^{t} \mathrm{e}^{A (\bkappa) (t-s)}{F}_i (\bkappa) \delta(s) \, \mathrm{d} s 
	\, = \, 
	C (\bkappa) \mathrm{e}^{A (\bkappa) t} {F}_i (\bkappa).
\end{equation}
Thus, the impulse response is directly obtained from the matrix exponential at a given time and the inverse Fourier transform in wall-parallel directions yields a solution in physical space.

	\vspace*{-2ex}
\subsection{Model parameters}
	\label{subsec:model_param}

The linearized equations \eqref{eq:6_lnsNV} contain four parameters: the solvent contribution to the shear viscosity, $\beta$, the amount of extensibility of the polymer molecules, $L$, the Weissenberg number, $\We$, and the Reynolds number, $Re$. A larger value of $\beta$ implies a smaller polymer concentration. A larger value of $\We$ implies that the fluid has a longer relaxation time. Due to the high computational expense of performing three-dimensional calculations on eight state variables (see \S~\ref{geef}), we restrict our analysis to a limited range of parameters; our choice represents a compromise between values used in experimental studies and the need to avoid numerical instabilities.

We present results for $Re = 50$, which is well within the laminar flow limits for a Newtonian fluid. In straight channels, elastic turbulence has been reported at smaller Reynolds numbers~\cite{qin2016elastic,pan2013nonlinear}; for example, recent experiments have shown turbulent features for Reynolds numbers between 2.5-150 (based on the half-channel width)~\cite{nolan2016viscoelastic}. We choose $Re = 50$ because resolving flow structures in physical space (\S~\ref{sec: Flow Structures}) at lower values of $Re$  requires larger values of $k_x$ and $k_z$, which in turn requires a larger number of discrete Fourier modes for good resolution in physical space. Flow structures presented in \S~\ref{sec: Flow Structures} use $512\times 512$ linearly spaced grid points in the $\bkappa$-plane with $\{k_{x,\min}=-50, k_{x,\max} =49.80\}$ and  $\{k_{z,\min}=-72, k_{z,\max} =71.72\}$. Larger values of $k_x$ and $k_z$ also require more Chebyshev collocation points to discretize the wall-normal direction (see \S~\ref{numer_meth}) sufficiently to avoid numerical instabilities \cite{graham1998effect,renardy1986linear}, which further increases the computational cost.


In \S~\ref{sec:Energy_Amplification} we present a parametric study for a range of Weissenberg numbers, but we choose a representative value of $\We= 50$ for most results presented here. This is because we found that the amplification increases with $\We$ (see \S~\ref{sec:Energy_Amplification}), and $\We = 50$ is the maximum value we could reach for grid-independent results without numerical instabilities. Confining ourselves to $\We\leq 50$ fixes the upper limit of the elasticity number $\mu = \We/Re $ in our simulations to $\mu = 1$. Using larger values of the Weissenberg number could bring out more distinctly features related to viscoelastic effects in flow structures, but at the cost of encountering and addressing numerical instabilities. In experiments concerning elastic turbulence, the Weissenberg number was varied between $20$ and $1000$~\cite{nolan2016viscoelastic}. We note that all results presented in this work are free from any artificial diffusion, numerical filters, or diffusion-inducing numerical schemes commonly employed to address numerical instabilities when simulating viscoelastic channel flows~\cite{lee_zaki_2017,PhysRevE.91.033013,HOUSIADAS2004243}. 

Groisman and Steinberg~\cite{groisman2000elastic} had $\beta = 0.765$ in curvilinear flows and refs.~\cite{pan2013nonlinear,qin2016elastic} had a value between $0.25$ and $0.5$ for straight-channel flows. Unless otherwise noted, we choose $\beta = 0.5$. In modeling polymeric fluids, values of $L$ have ranged widely, from about $2.5$ to infinity~\cite{chilcott1988creeping}. In the limit of infinite $L$, the FENE-CR model reduces to the Oldroyd-B model, which is successful in describing some features of dilute polymeric flows but becomes less accurate at higher shear rates \cite{larson1999structure,bird1987dynamics}. Since finite values of $L$ have been shown to work better in modeling dilute polymeric flows~\cite{larson1999structure,bird1987dynamics,chilcott1988creeping}, we set $L = 100$.


	\vspace*{-2ex}
\subsection{Energy of velocity fluctuations}\label{KE}

The integral of the kinetic energy of the velocity fluctuations in the wall-normal direction can be evaluated using a weighted inner product of the output with itself
\begin{equation}\label{eq:9}
	E_{i}(\bkappa,t)
	\; \DefinedAs \;
  	\int_{-1}^1
	\BB{v}_i^* (\bkappa,y,t) \, \BB{v}_i (\bkappa,y,t)  
	\, 
	\mathrm{d}y
	\; = \;
	\phi_{i}^* (\bkappa,t) \, I_w \, \phi_{i} (\bkappa,t),
\end{equation}
where $(\cdot)^*$ denotes the complex conjugate transpose and $I_w$ is a diagonal matrix of the appropriate integration weights for the Chebyshev collocation points. We recall that the subscript $i$ denotes the input direction of the impulsive excitation, cf.~\eqref{16b}.

We further perform integration over time to obtain
\begin{subequations}\label{eq:10}
	\begin{equation}
	\label{eq:10a}
	 \bar{E}_{i}(\bkappa) 
	 \; \DefinedAs \;
	\int_{0}^{\infty}
	 {E}_{i}(\bkappa,t) 
	 \,
	 \mathrm{d}t,
\end{equation}
and note that for stable systems, the solution to the algebraic Lyapunov equation \cite{jovbamJFM05},
\begin{equation}\label{eq:10b}
A(\bkappa)X_{i}(\bkappa) + X_{i}(\bkappa)A^\dagger(\bkappa)\, =\, -F_i(\bkappa)F_i^\dagger(\bkappa),\\
\end{equation}
can be used to avoid explicit integration in~\eqref{eq:10a} and compute $\bar{E}_{i}(\bkappa)$ as
	\begin{equation}\label{eq:10c}
	\bar{E}_{i}(\bkappa)
	\; = \;
	\trace 
	\, 
	(X_{i}(\bkappa) \, C^\dagger (\bkappa) \, C(\bkappa)).
	\end{equation}
	\end{subequations}
Here, $(\cdot)^\dagger$ is the finite-dimensional approximation to the adjoints of the operators that appear in~\eqref{eq:11} and $\trace \, (\cdot)$ is the matrix trace, i.e., the sum of its eigenvalues. Adjoints are defined with respect to a weighted inner product that determines the kinetic energy of velocity fluctuations~\cite{jovbamJFM05,butler1992three}; see Appendix~\ref{appB} for additional details. 

In addition to the total kinetic energy, we also analyze the componentwise contribution of velocities $r = u$, $v$, or $w$ to the total kinetic energy,
	\begin{subequations}
	\begin{align}
	\label{eira}
	E_{ri}(\bkappa,t) 
	& \; \DefinedAs \; 
	\int_{-1}^1
	r_i^* (\bkappa,y,t) \, r_i (\bkappa,y,t) \, \mathrm{d}y,
	\\[0.15cm]
	\label{eirb}
	\bar{E}_{ri}(\bkappa) 
	& \; \DefinedAs \;
	\int_{0}^{\infty}
	 {E}_{ri}(\bkappa,t) 
	 \,
	 \mathrm{d}t,
	\end{align}
	\end{subequations}
where $E_{ri}$ and $\bar{E}_{ri}$ represent the energy of the velocity component $r$ arising from the impulsive forcing in the $i$th coordinate direction. We note that 
	\[
	E_{i} 
	\; = \; 
	E_{ui} 
	\; + \; 
	E_{vi} 
	\;+ \;
	E_{wi},
	\] 
and that $E_{ri}$ and $\bar{E}_{ri}$ can be evaluated in a similar manner as the total kinetic energy in~\eqref{eq:9} and~\eqref{eq:10} by replacing $C(\bkappa)$ in~\eqref{eq:10c} with $C_r(\bkappa)$; see equation~\eqref{apC} in Appendix~\ref{appA}.

	\vspace*{-2ex}
\section{Flow sensitivity to the location of the impulse}\label{sec:y0loc}
The location of the impulse in the ($x,z$)-plane is immaterial because the $x$- and $z$-directions are translationally invariant. However, the sensitivity of the flow may vary with the choice of the location of the impulse in the wall-normal direction. We next examine how the sensitivity of a viscoelastic channel flow changes with the wall-normal location of impulsive forcing.  

  \begin{figure}
  \centering
  \subfloat[][]{\includegraphics[scale=0.36]{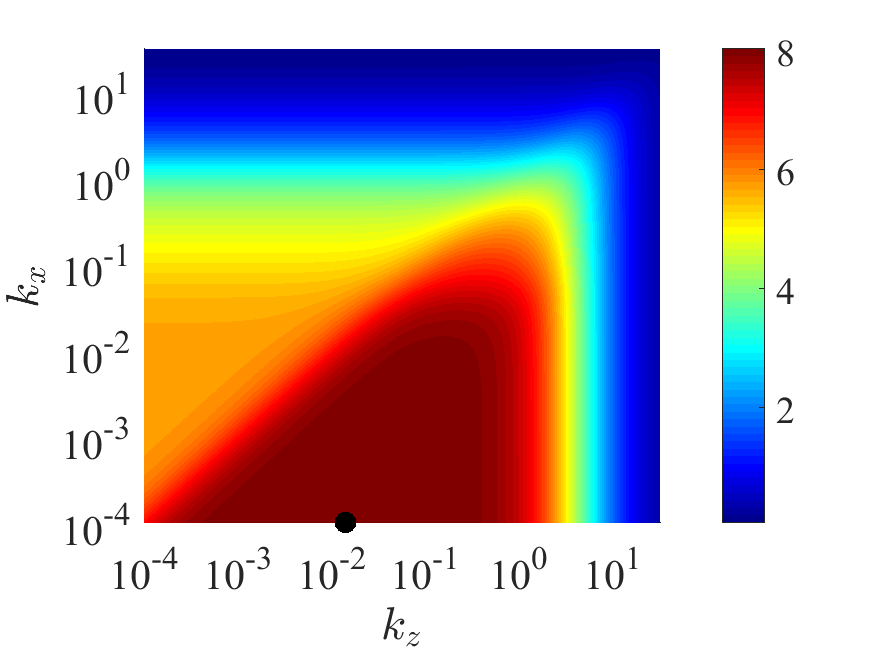}\label{fig:2a} }
  \subfloat[][]{\includegraphics[scale=0.36]{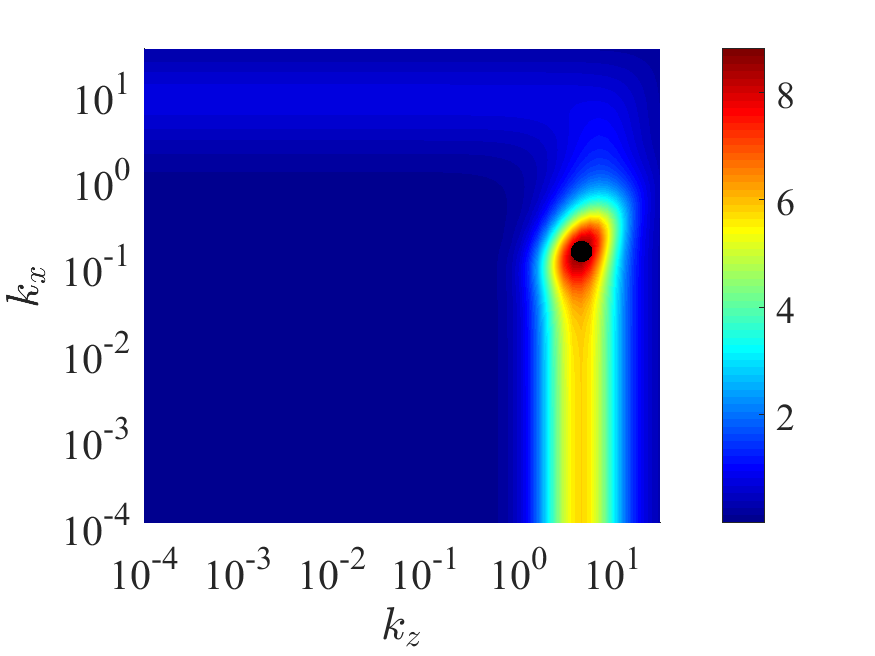}\label{fig:2b}}
  \subfloat[][]{\includegraphics[scale=0.36]{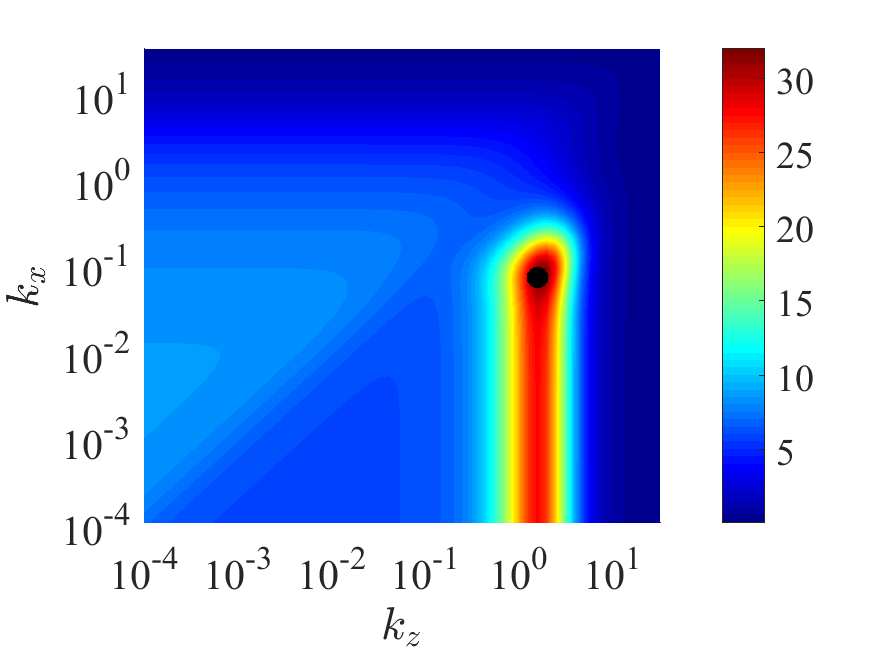}\label{fig:2c}}
  \caption{\label{fig:2} Kinetic energy integrated over the wall-normal direction and time, with an impulsive excitation located at $y_0 = -0.75$. Plots correspond to  \protect\subref{fig:2a} $\bar{E}_x$, \protect\subref{fig:2b} $\bar{E}_y$, and \protect\subref{fig:2c} $\bar{E}_z$ respectively, calculated from \eqref{eq:10}. Parameters used are $Re = 50$, $\We = 50$, $L = 100$, and $\beta = 0.5$. The maximum value of the kinetic energy is marked by the black dots.}
  \end{figure}

\begin{figure}
  \centering
  \subfloat[][]{\includegraphics[scale=0.36]{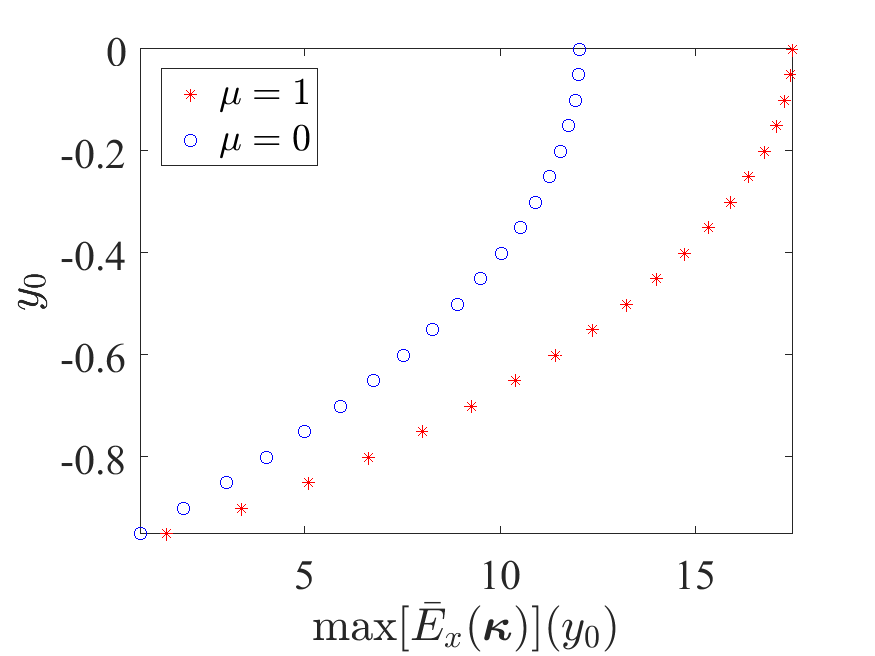}\label{fig:4a} }
  \subfloat[][]{\includegraphics[scale=0.36]{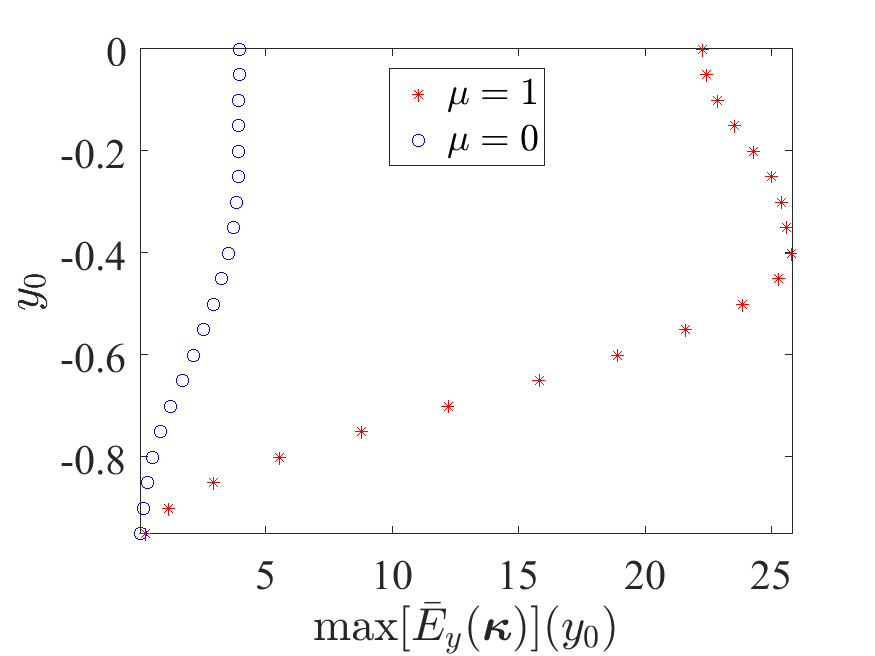}\label{fig:4b}}
  \subfloat[][]{\includegraphics[scale=0.36]{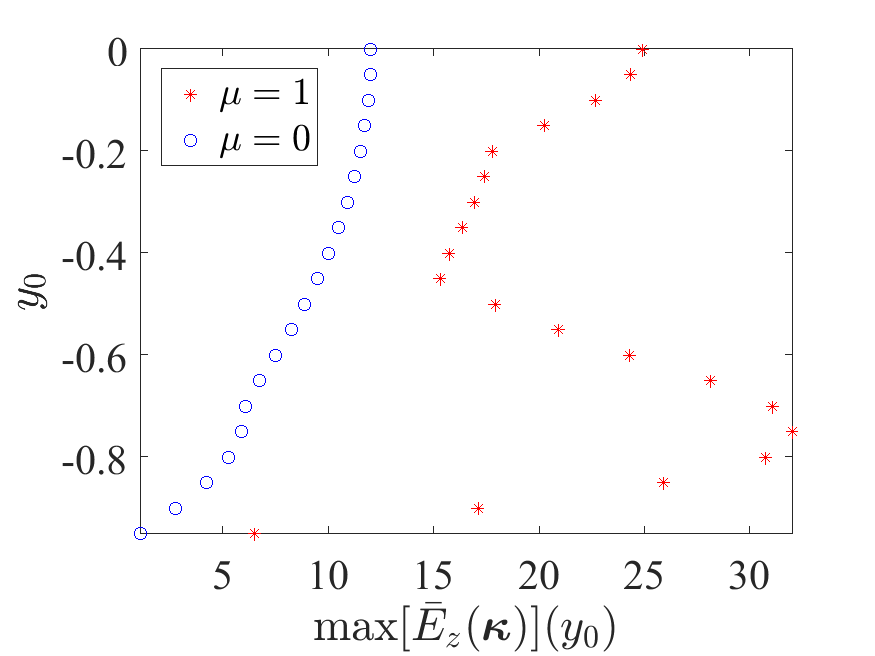}\label{fig:4c}}
  \caption{\label{fig:4} Maximum kinetic energy induced by an impulsive excitation in the \protect\subref{fig:4a} streamwise, \protect\subref{fig:4b} wall-normal, and \protect\subref{fig:4c} spanwise directions as a function of $y_0$, calculated using \eqref{eq:10}. The Newtonian fluid corresponds to $\mu = 0$, and the viscoelastic fluid to $\mu = 1$ (the parameters are $Re = 50$, $\We = 50$, $L = 100$, and $\beta = 0.5$).}
  \end{figure}

Due to the symmetry of plane Poiseuille flow, we only consider the lower half of the channel. We first calculate the kinetic energy averaged over the wall-normal direction and time (using~\eqref{eq:10}) as a function of $\bkappa$ for different values of $y_0$. For example, Figure \ref{fig:2} shows the kinetic energy for an impulsive excitation in the streamwise direction (Figure \ref{fig:2a}), wall-normal direction (Figure \ref{fig:2b}), and spanwise direction (Figure \ref{fig:2c}) for $y_0 = -0.75$ and $Re = 50$. The maximum value of the kinetic energy over all values of $k_x$ and $k_z$ is marked by the black dots in Figure \ref{fig:2}. In Figure \ref{fig:4}, we examine how these peak values depend on the location of the impulse, $y_0$. We note that for $y_0 = -0.75$ the maximum value of the kinetic energy  in Figure \ref{fig:2a} occurs at ($k_x \approx 10^{-4}\text{, }k_z\approx10^{-2}$), in Figure \ref{fig:2b} at ($k_x \approx 10^{-1}\text{, }k_z\approx 1.5$), and in Figure \ref{fig:2c} at ($k_x \approx 10^{-1}\text{, }k_z\approx 10^0$). These values change as we change $y_0$. 

Figure \ref{fig:4} shows how the largest value of kinetic energy depends on the wall-normal location $y_0$ of an impulsive excitation in the streamwise (Figure \ref{fig:4a}), wall-normal (Figure \ref{fig:4b}), and spanwise (Figure \ref{fig:4c}) direction. The relative contribution of fluid elasticity compared to vorticity diffusion can be quantified in terms of the elasticity number $\mu= \We/Re$. In Figure \ref{fig:4}, the Newtonian fluid corresponds to $\mu = 0$ and the viscoelastic fluid to $\mu = 1$; as mentioned in \S~\ref{subsec:model_param}, since we set $Re = 50$ and confine our attention to $\We\leq 50$, in our study we have $0 \leq \mu \leq 1$. 
 
We see that the introduction of viscoelasticity increases the kinetic energy of velocity fluctuations for an impulsive excitation in any of the three directions. Larger amplification of disturbances in viscoelastic fluids indicates their greater sensitivity at relatively low values of the Reynolds number. As Figure \ref{fig:4a} demonstrates, the influence of viscoelasticity is not as significant for an impulsive excitation in the streamwise direction. We observe more pronounced differences between Newtonian and viscoelastic responses for excitations in the wall-normal (Figure \ref{fig:4b}) and spanwise (Figure \ref{fig:4c}) directions. The largest discrepancy between corresponding kinetic energies occurs for the impulse in the spanwise direction (Figure \ref{fig:4c}) at a location $y_0 = -0.75$.

\begin{figure}
  \centering
  \subfloat[][]{\includegraphics[scale=0.36]{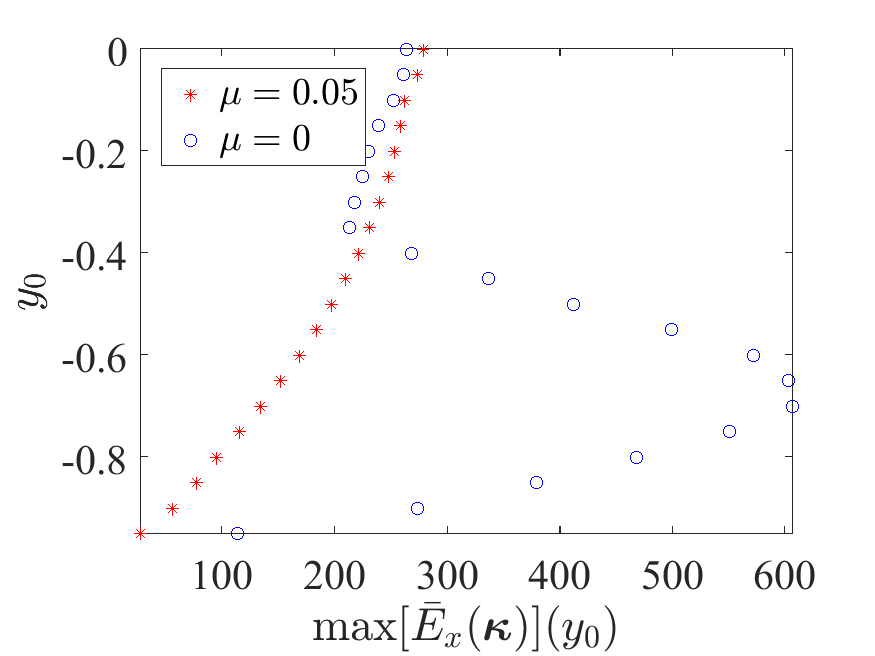}\label{fig:5a} }
  \subfloat[][]{\includegraphics[scale=0.36]{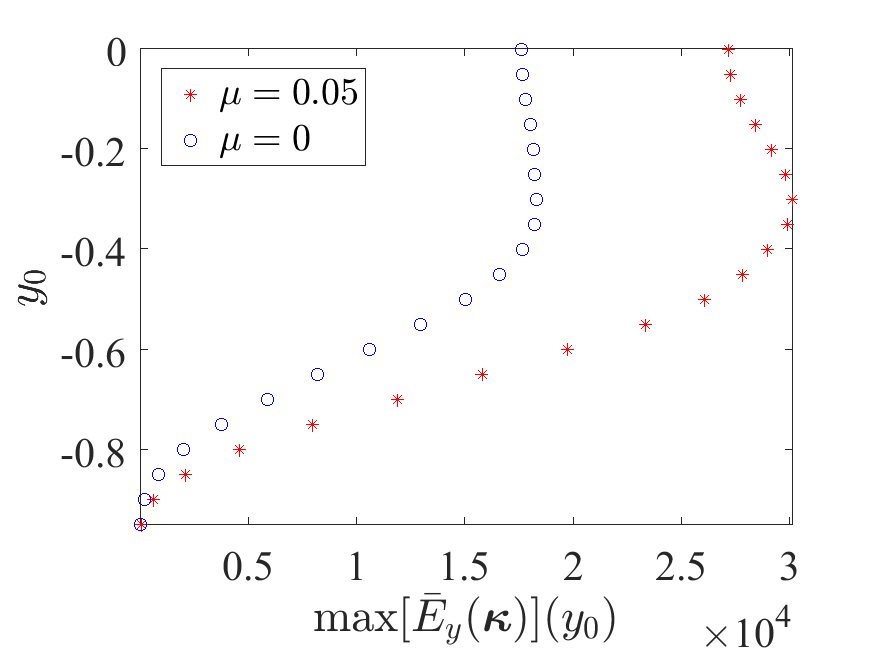}\label{fig:5b}}
  \subfloat[][]{\includegraphics[scale=0.36]{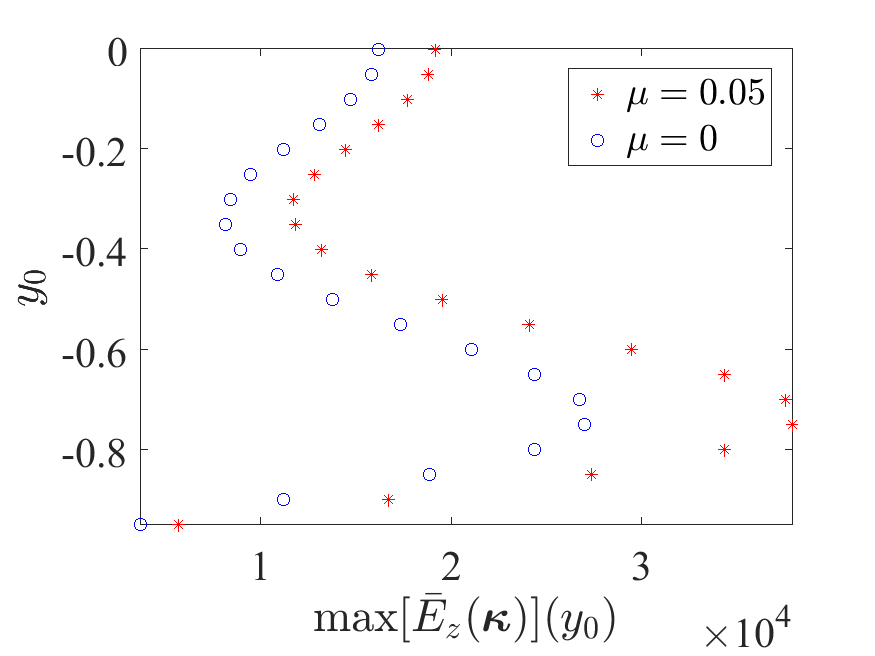}\label{fig:5c}}
\caption{\label{fig:5} Maximum energy induced by an impulsive excitation in the \protect\subref{fig:5a} streamwise, \protect\subref{fig:5b} wall-normal, and \protect\subref{fig:5c} spanwise directions as a function of $y_0$, calculated using \eqref{eq:10}. The Newtonian fluid corresponds to $\mu = 0$, and the viscoelastic fluid to $\mu = 0.05$ (the parameters are $Re = 1000$, $\We = 50$, $L = 100$, and $\beta = 0.5$).}
  \end{figure}

Thus far we have studied the sensitivity of Newtonian and viscoelastic plane Poiseuille flow to the wall-normal location of point forces when $Re = 50$. Increasing $Re$ to larger values shows interesting similarities between a Newtonian fluid at high Reynolds numbers and a viscoelastic fluid at low Reynolds numbers. Figure \ref{fig:5} shows similar plots of the maximum kinetic energy over all $k_x$ and $k_z$ when $Re = 1000$. We see that the plots for the largest kinetic energy for impulsive excitations in the wall-normal (Figure \ref{fig:5b}) and spanwise (Figure \ref{fig:5c}) directions are similar in shape for the Newtonian and viscoelastic fluids. This is because inertial forces dominate over elastic forces as reflected by the value of the elasticity numbers, $\mu = 0.05$ (viscoelastic) and $\mu = 0$ (Newtonian). The plots have very different shapes for a Newtonian and a viscoelastic fluid with an impulsive streamwise excitation (Figure \ref{fig:5a}), in that the viscoelastic fluid is less energetic at high Reynolds numbers. However, the values of energy (as seen from the $y$-axis) are substantially lower when compared to impulsive excitations in the spanwise or wall-normal directions.

It is interesting to observe a similar maximum near the wall in Figure \ref{fig:5c} in the Newtonian fluid at $Re=1000$ that was seen in the viscoelastic fluid at $Re = 50$ (the peak located at $y_0 = -0.75$ in Figure \ref{fig:4c}). We note that this maximum was absent in the Newtonian fluid at $Re = 50$ (Figure \ref{fig:4c}). This indicates a striking similarity in the nonmodal amplification of a Newtonian fluid at high Reynolds numbers and a viscoelastic fluid at low Reynolds numbers and can be attributed to the viscoelastic analogue of the well-known lift-up mechanism~\cite{jovkumPOF10,jovkumJNNFM11}. 

%

The governing mechanism in Newtonian fluids that leads to nonmodal amplification comes from the vortex-tilting effect, and can be analyzed by examining the equation for the evolution of the wall-normal vorticity $\eta$, 
\begin{equation}\label{vortex_tilting_N}
\partial_t\eta \; = \; -Re\, U'\,\partial_z v \; + \; \Delta\eta,
\end{equation}
where the time is scaled with the diffusive time scale $h^2 \rho/\eta_T$. The second term in \eqref{vortex_tilting_N} is the Laplacian operator $\Delta$ that arises from viscous dissipation and it acts to decrease the wall-normal vorticity. The first term is solely responsible for non-modal amplification. The term $-U'$ is the spanwise vorticity in the base flow, and $\partial_zv$ can be interpreted as a stress which corresponds to a force in the wall-normal direction that varies in the spanwise direction. Thus the spanwise vorticity in the base flow is forced in the wall-normal direction so as to amplify the wall-normal vorticity \cite{butler1992three}.

For inertialess Oldroyd-B fluids, Jovanovi\'c and Kumar~\cite{jovkumJNNFM11} derived the evolution equation for the wall-normal vorticity $\eta$ by scaling the time with the relaxation time of the polymer, 
\begin{equation}\label{vortex_tilting_P}
\partial_t \, \Delta \eta  
	\; = \; 
	-\We\, \frac{1 - \beta}{\beta}
	\, (U'(y)\Delta\partial_z \, + \, 2U''(y)\partial_{yz}) \, v 
	\, - \, 
	\frac{1}{\beta} \, \Delta \eta.
\end{equation}
For Newtonian fluids, the vortex tilting term in \eqref{vortex_tilting_N} vanishes in the absence of inertia, i.e., at $Re = 0$. Viscoelastic fluids, however, have additional terms that can produce a vortex-tilting-like effect even at $Re=0$; cf.~\eqref{vortex_tilting_P}. The terms $U'$ and $U''$ in \eqref{vortex_tilting_P} come from the shear stress $T_{12}$ in the base flow of the viscoelastic fluid~\cite{jovkumJNNFM11}.

Equations \eqref{vortex_tilting_N} and \eqref{vortex_tilting_P} suggest an underlying similarity between a Newtonian fluid at high Reynolds numbers and a viscoelastic fluid at low Reynolds numbers. A Newtonian fluid at high Reynolds numbers experiences amplification due to the spanwise vorticity in the base state. Similarly, a viscoelastic fluid at low Reynolds numbers experiences amplification due to a coupling between the polymeric stresses in the base state and velocity fluctuations. The amount of amplification scales with the Weissenberg number for the inertialess viscoelastic fluid and with the Reynolds number for the Newtonian fluid.

In this section, we have investigated the influence of the location of the impulse on the flow. We have demonstrated that the impulse in the spanwise direction has the maximum impact on the flow and that there is similarity in the nature of the most sensitive locations between a Newtonian fluid at high Reynolds numbers and a viscoelastic fluid at low Reynolds numbers. This similarity can be understood in terms of the well-known lift-up mechanism and its viscoelastic analogue, as  discussed by Jovanovi\'c and Kumar~\cite{jovkumJNNFM11}. In \S~\ref{sec:Energy_Amplification}, we examine energy of velocity fluctuations corresponding to an impulsive excitation at $y_0 = -0.75$ in a flow with $Re = 50$.

	\vspace*{-2ex}
\section{Energy evolution}\label{sec:Energy_Amplification}

In the previous section, we identified the location in the wall-normal direction where the localized point force has the maximum impact on the flow. In this section, we contrast the evolution of energy in viscoelastic and Newtonian fluids by introducing an impulse at the optimal location for viscoelastic fluids. We examine the impact on the streamwise, wall-normal, and spanwise velocity fluctuations separately and find that the streamwise velocity is most affected. We then study changes in the energy evolution with $\beta$ (polymer concentration) and $\We$ (polymer relaxation time).

\begin{figure}
\centering
$\beta = 0.5$, $L = 100$, $\mu = 1$, and $Re = 50$ (Viscoelastic fluid):\\
\subfloat[][$\bar{E}_x$]{\includegraphics[scale=0.36]{huvw_x}\label{fig:7a} }
\subfloat[][$\bar{E}_y$]{\includegraphics[scale=0.36]{huvw_y}\label{fig:7b}}
\subfloat[][$\bar{E}_z$]{\includegraphics[scale=0.36]{huvw_z}\label{fig:7c}}\\
\vspace{0.25cm}
$\mu = 0$ and $Re = 50$ (Newtonian fluid):\\
\subfloat[][$\bar{E}_x$]{\includegraphics[scale=0.36]{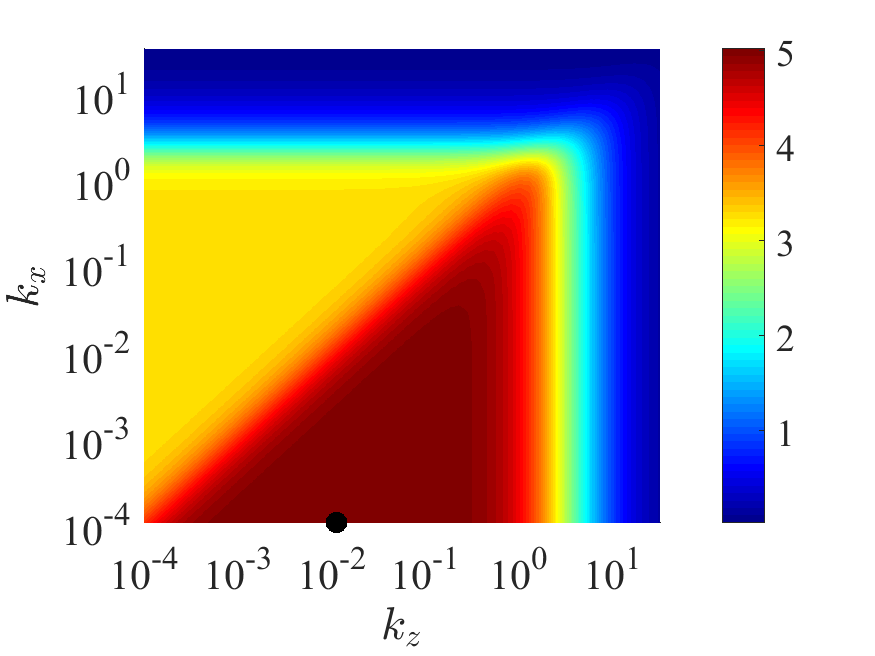}\label{fig:7d} }
\subfloat[][$\bar{E}_y$]{\includegraphics[scale=0.36]{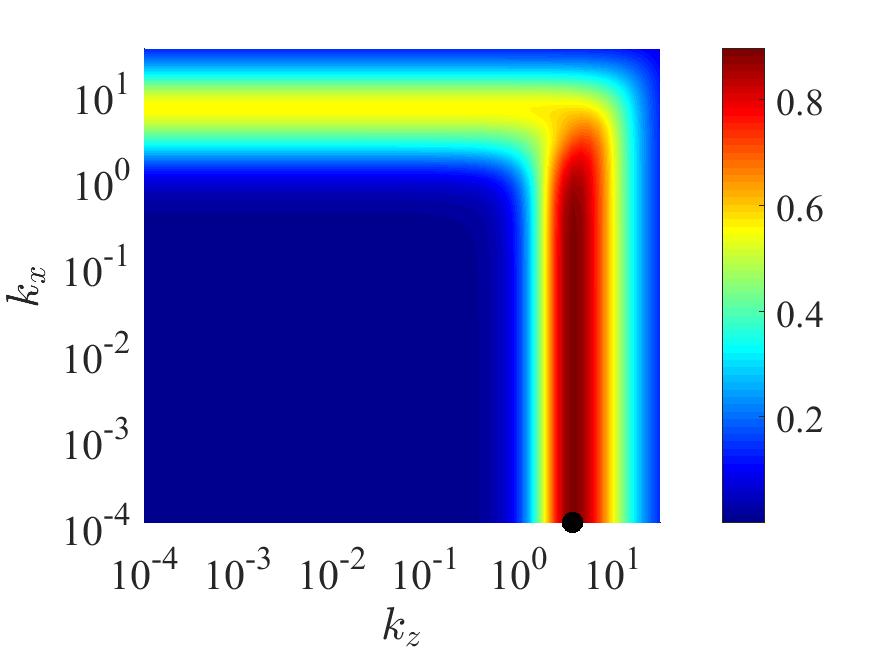}\label{fig:7e}}
\subfloat[][$\bar{E}_z$]{\includegraphics[scale=0.36]{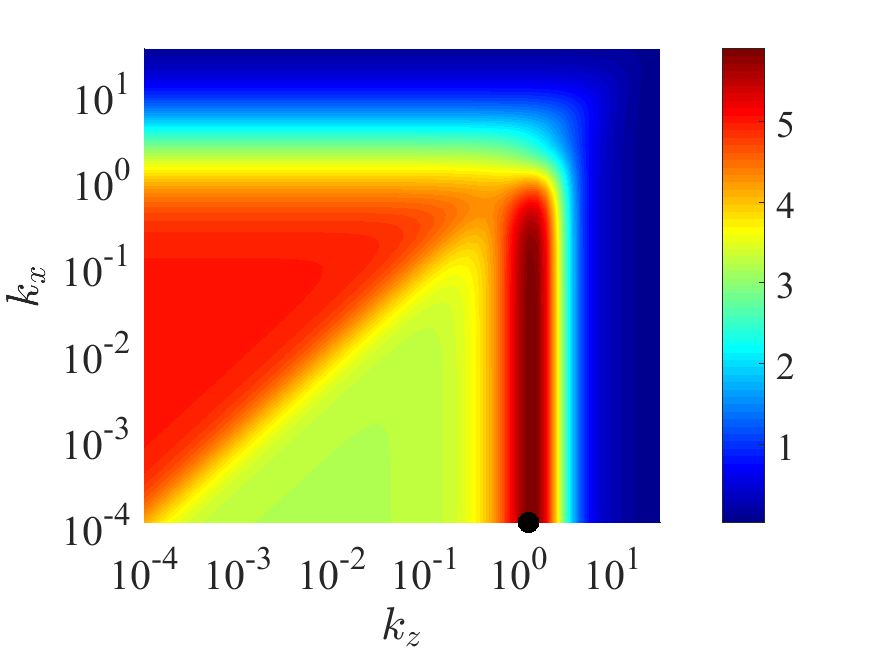}\label{fig:7f}}\\
\caption{\label{fig:7} Kinetic energy integrated over the wall-normal direction and time, with an impulse in the direction $i$, calculated using \eqref{eq:10}. We consider a viscoelastic fluid with an impulsive excitation in the \protect\subref{fig:7a} streamwise, \protect\subref{fig:7b} wall-normal, and \protect\subref{fig:7c} spanwise directions, and a Newtonian fluid with an impulsive excitation in the \protect\subref{fig:7d} streamwise, \protect\subref{fig:7e} wall-normal, and \protect\subref{fig:7f} spanwise directions. The maximum value of the kinetic energy is marked by the black dots.} 
\end{figure}

Figure \ref{fig:7} shows the kinetic energy averaged over the wall-normal direction and time at $Re = 50$, calculated from \eqref{eq:10}. Figures \ref{fig:7a}-\ref{fig:7c} show the kinetic energy of a viscoelastic fluid and Figures \ref{fig:7d}-\ref{fig:7f} show the kinetic energy of a Newtonian fluid. Figures \ref{fig:7a} and \ref{fig:7d} correspond to an impulsive forcing in the streamwise direction, Figures \ref{fig:7b} and \ref{fig:7e} correspond to an impulsive excitation in the wall-normal direction, and Figures \ref{fig:7c} and \ref{fig:7f} correspond to an impulsive excitation in the spanwise direction. By observing the scales on the color bars we see that, in all cases, the kinetic energy of velocity fluctuations in the viscoelastic fluid is higher when compared to a Newtonian fluid. The additional energy in viscoelastic fluids comes from the elastic stresses in the base flow which are absent in Newtonian fluids \cite{jovkumJNNFM11,liejovkumJFM13}; see the discussion toward the end of \S~\ref{sec:y0loc} (and \eqref{vortex_tilting_P}). 

It can be seen that there is not a significant difference in the kinetic energy for a streamwise impulsive excitation in the viscoelastic (Figure \ref{fig:7a}) and Newtonian (Figure \ref{fig:7d}) fluids. For impulsive excitations in the wall-normal and spanwise directions, however, differences are significant. Impulsive excitations in the wall-normal and spanwise directions for the viscoelastic fluid produce fluctuations that are nearly streamwise-constant with the maximum kinetic energy near $k_x \approx 10^{-1},k_z\approx 10^0$ (Figures \ref{fig:7b} and \ref{fig:7c}). For the Newtonian fluid, the resulting fluctuations are less oblique ($k_x \approx 10^{-4}$; Figures \ref{fig:7e} and \ref{fig:7f}).

The spanwise impulsive excitation is amplified about six times more in viscoelastic fluid (Figure \ref{fig:7c}) than in the Newtonian fluid (Figure \ref{fig:7f}). Since the impulse in the spanwise direction induces the highest amount of energy, in what follows we only analyze the impact of the spanwise impulsive excitation on the evolution of velocity fluctuations.

\begin{figure}
\centering
\subfloat[][$\bar{E}_{uz}$]{\includegraphics[scale = 0.36]{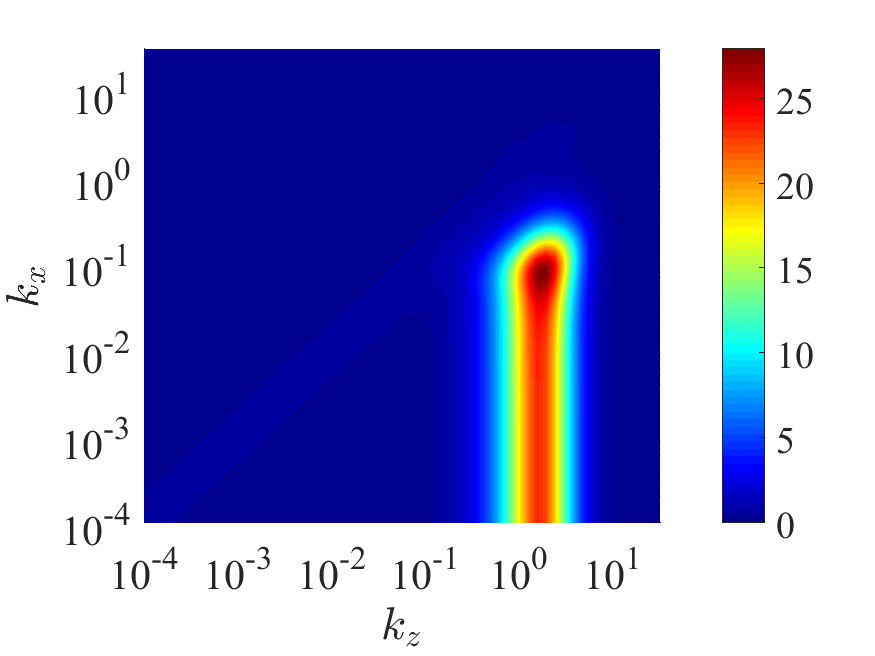}\label{fig:Guvw_a}}
\subfloat[][$\bar{E}_{vz}$]{\includegraphics[scale = 0.36]{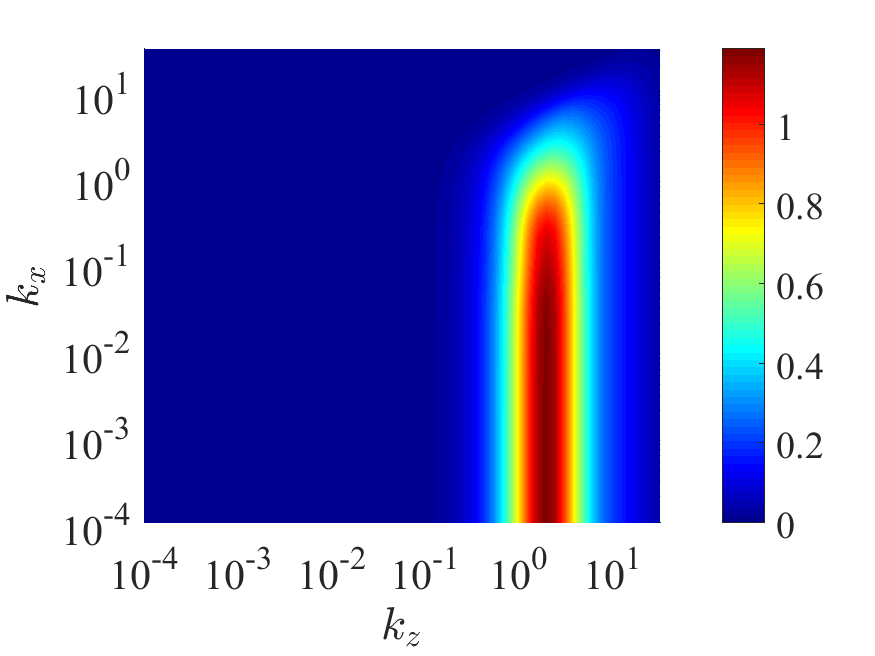}\label{fig:Guvw_b}}
\subfloat[][$\bar{E}_{wz}$]{\includegraphics[scale = 0.36]{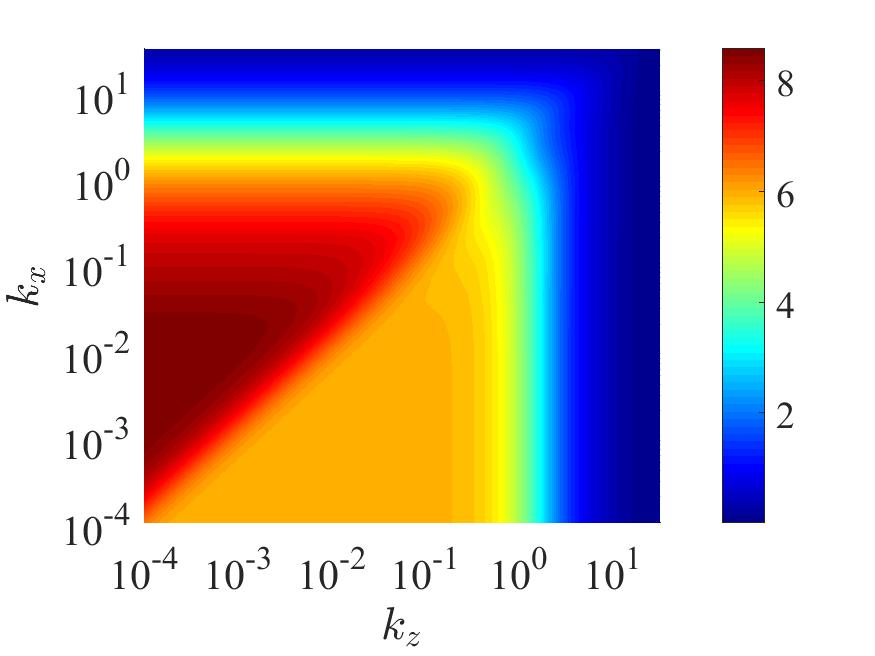}\label{fig:Guvw_c}}
\caption{\label{fig:Guvw}Componentwise contributions of \protect\subref{fig:Guvw_a} streamwise, \protect\subref{fig:Guvw_b} wall-normal, and \protect\subref{fig:Guvw_c} spanwise velocities to the total kinetic energy calculated from \eqref{eirb} arising from an impulsive spanwise forcing in a flow with $Re = 50$, $\We = 50$, $L=100$, and $\beta = 0.5$.}
\end{figure}

The energy can be further analyzed based on the individual contributions from the streamwise, wall-normal, and spanwise velocities. Figure \ref{fig:Guvw} shows the contribution of the total kinetic energy due to the  streamwise velocity (Figure \ref{fig:Guvw_a}),  wall-normal velocity (Figure \ref{fig:Guvw_b}), and spanwise velocity (Figure \ref{fig:Guvw_c}). From the color bars, we notice that the streamwise velocity has the largest contribution to the overall energy. We thus conclude that the spanwise forcing has the maximum impact on the flow and that the streamwise velocity is most affected. This observation is again similar to what is seen for Newtonian fluids at high Reynolds numbers as investigated by Jovanovi\'c and Bamieh \cite{jovbamJFM05}. The difference is that the most amplified disturbances are more oblique ($k_x\approx 10^{-1}\text{, }k_z \approx 10^0$) in  viscoelastic fluids when compared to Newtonian fluids at high Reynolds numbers, where the most prominent fluctuations are streamwise-constant ($k_x\approx 0$, $k_z \approx 10^0$). The analysis presented in this section provides deeper insight into the individual energies of each velocity component.

\begin{figure}
\centering
\subfloat[][$Re = 50,\text{ }L = 100, \text{ }\beta = 0.5$]{\includegraphics[scale = 0.45]{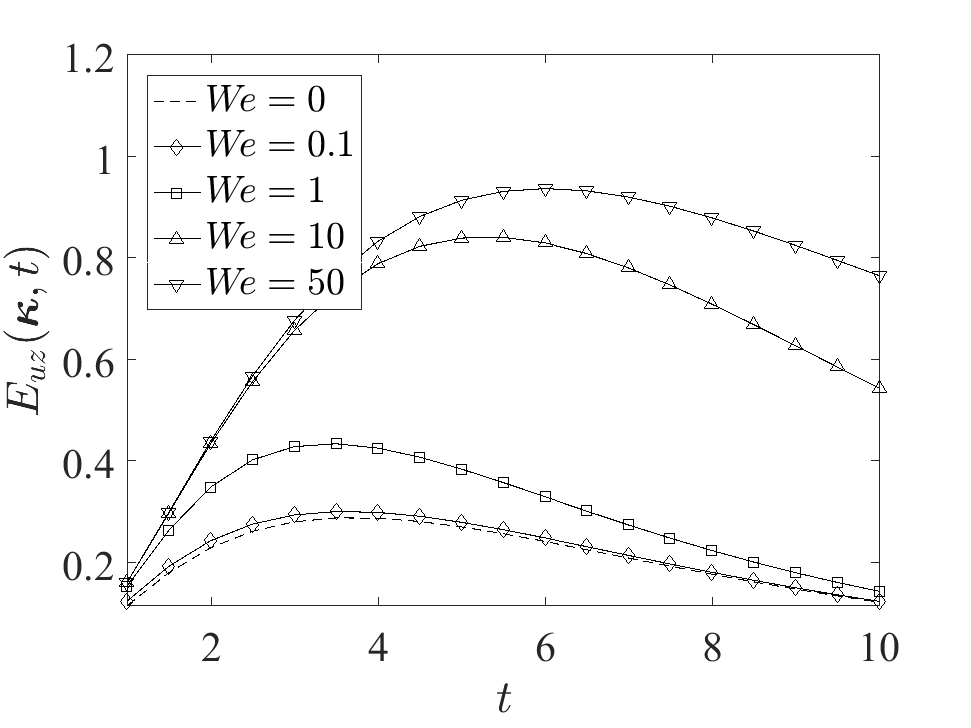}\label{fig:We}}
\subfloat[][$Re = 50,\text{ }L = 100, \text{ }\We = 50$]{\includegraphics[scale = 0.45]{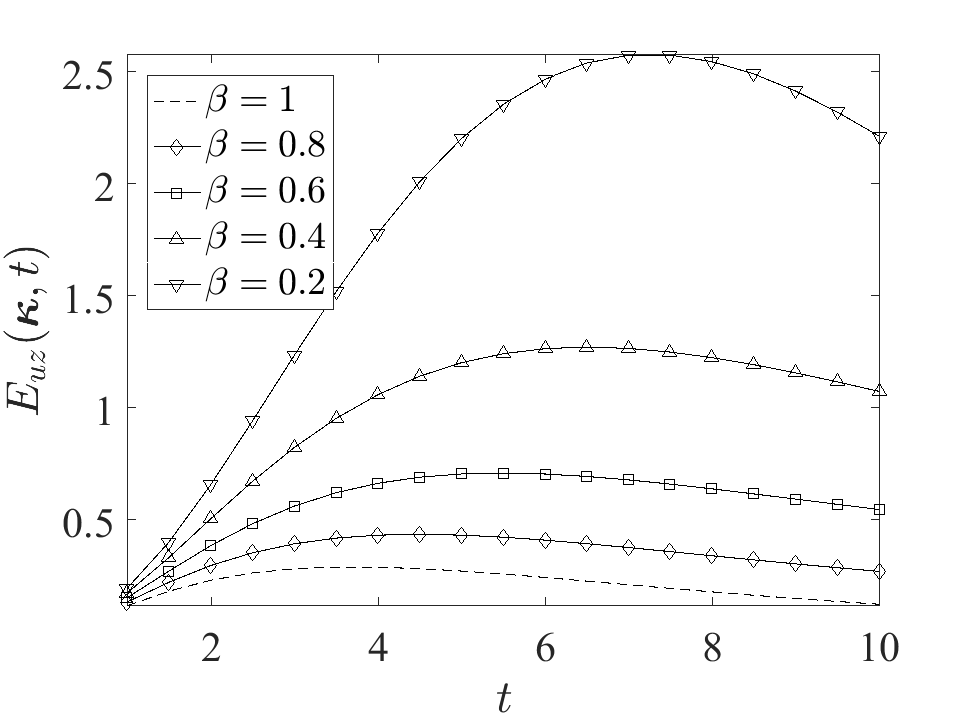}\label{fig:beta}}
\caption{\label{fig:Webeta}Transient evolution of kinetic energy of streamwise velocity fluctuations arising from an an impulse in the spanwise direction, $E_{uz}(\bkappa,t)$ calculated from \eqref{eira}, in a flow with $k_x = 10^{-1}$ and $k_z = 10^0$. Panel \protect\subref{fig:We} shows the effect of increasing the polymer relaxation time by increasing the Weissenberg number. $\We = 0$ corresponds to the Newtonian fluid. Panel \protect\subref{fig:beta} shows the effect of increasing polymer concentration, $1-\beta$. For a Newtonian fluid, $\beta = 1$.}
\end{figure}
Figure \ref{fig:Webeta} shows the transient evolution of the kinetic energy of the streamwise velocity fluctuation as a function of time; computations are done using~\eqref{eira}. Since Figure \ref{fig:Guvw} demonstrates that the streamwise velocity is most amplified, we plot only the energy of the streamwise velocity fluctuations with $k_x = 10^{-1}$ and $k_z = 10^0$, a wavenumber pair which corresponds to the black dot in Figure \ref{fig:7c}. In Figure \ref{fig:We}, we see that kinetic energy increases with increasing the Weissenberg number. We also see in Figure \ref{fig:beta} an increase in energy upon increasing the polymer concentration.

Experiments on microchannel flows of viscoelastic solutions with an induced disturbance in the form of a cylindrical obstruction were recently reported in ref.~\cite{nolan2016viscoelastic}. The Reynolds numbers were between $2.5$ and $150$. Flow instabilities were observed at a localized region in the vicinity of the obstruction which became more prominent with an increase in polymer concentration. Although we confine our attention to the class of impulsive excitations (which are different from the excitations considered in ref.~\cite{nolan2016viscoelastic}), we observe qualitative agreement in the sense that the transient energy amplification increases as we increase the polymer concentration. 

	\vspace*{-2ex}
\section{Spatio-temporal evolution of flow structures}\label{sec: Flow Structures}

\begin{figure}
\centering
\subfloat[][$t = 0.1$, Viscoelastic]{\includegraphics[scale = 0.36]{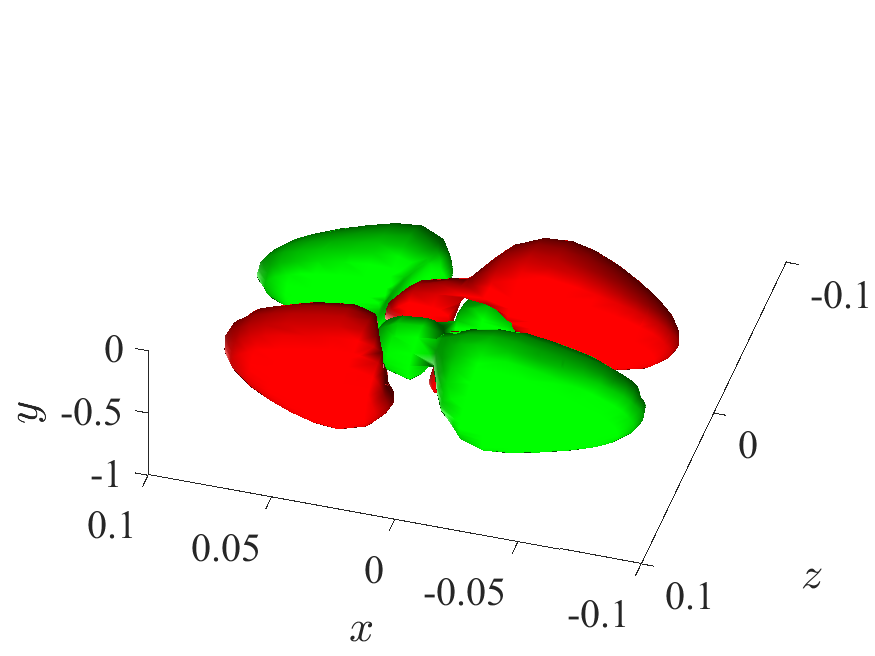}\label{fig:Isoz_a}}\qquad
\subfloat[][$t = 0.1$, Newtonian]{\includegraphics[scale = 0.36]{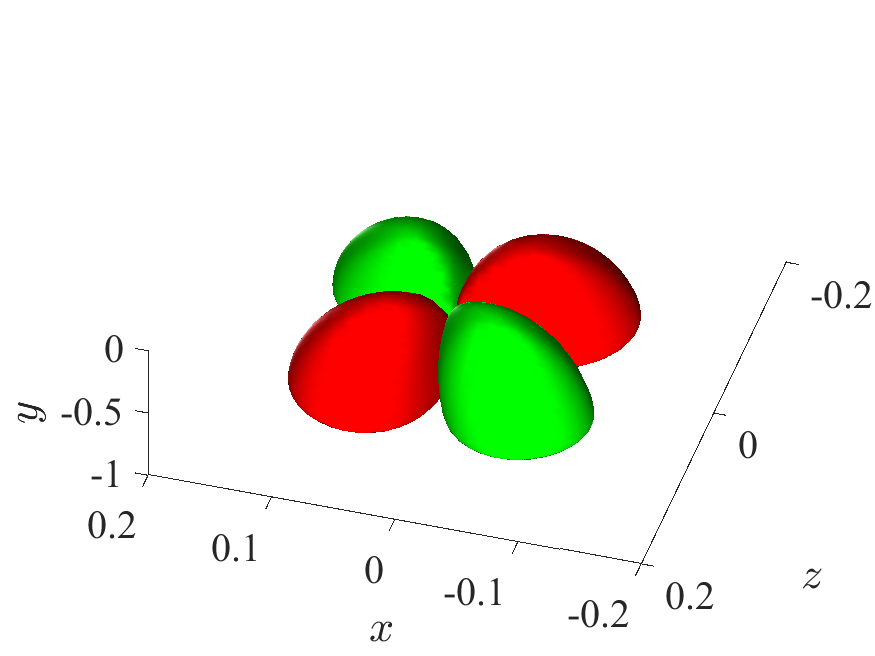}\label{fig:IsozN_a}}\\
\subfloat[][$t = 1$, Viscoelastic]{\includegraphics[scale = 0.36]{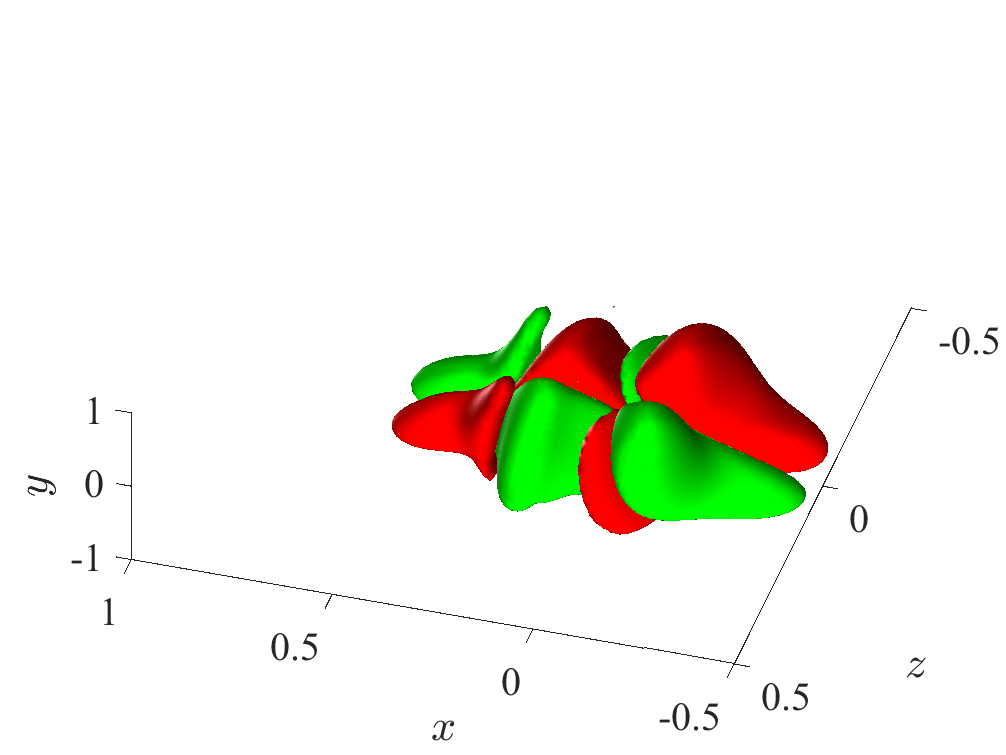}\label{fig:Isoz_b}}\qquad
\subfloat[][$t = 1$, Newtonian]{\includegraphics[scale = 0.36]{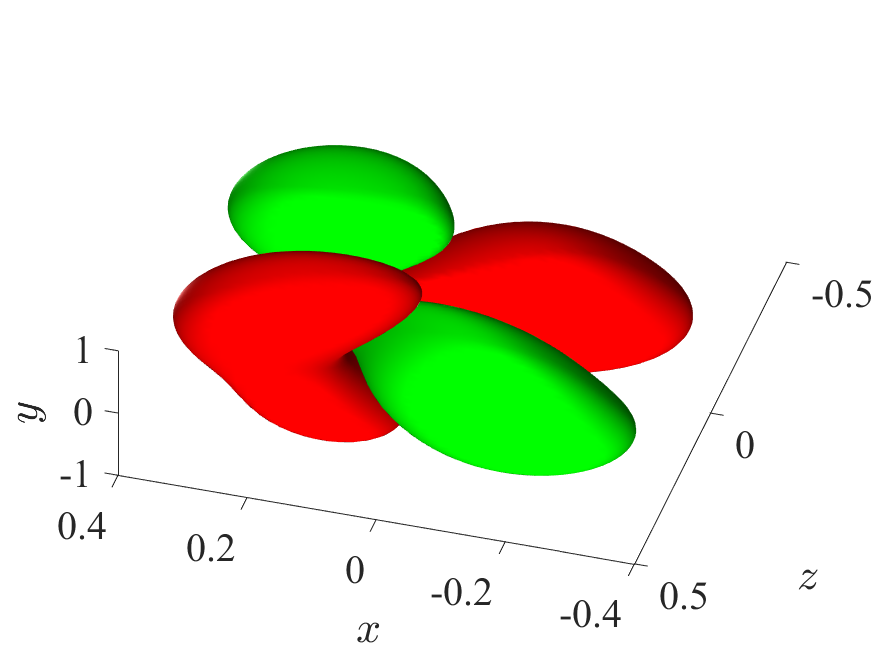}\label{fig:IsozN_b}}\\
\subfloat[][$t = 3.5$, Viscoelastic]{\includegraphics[scale = 0.36]{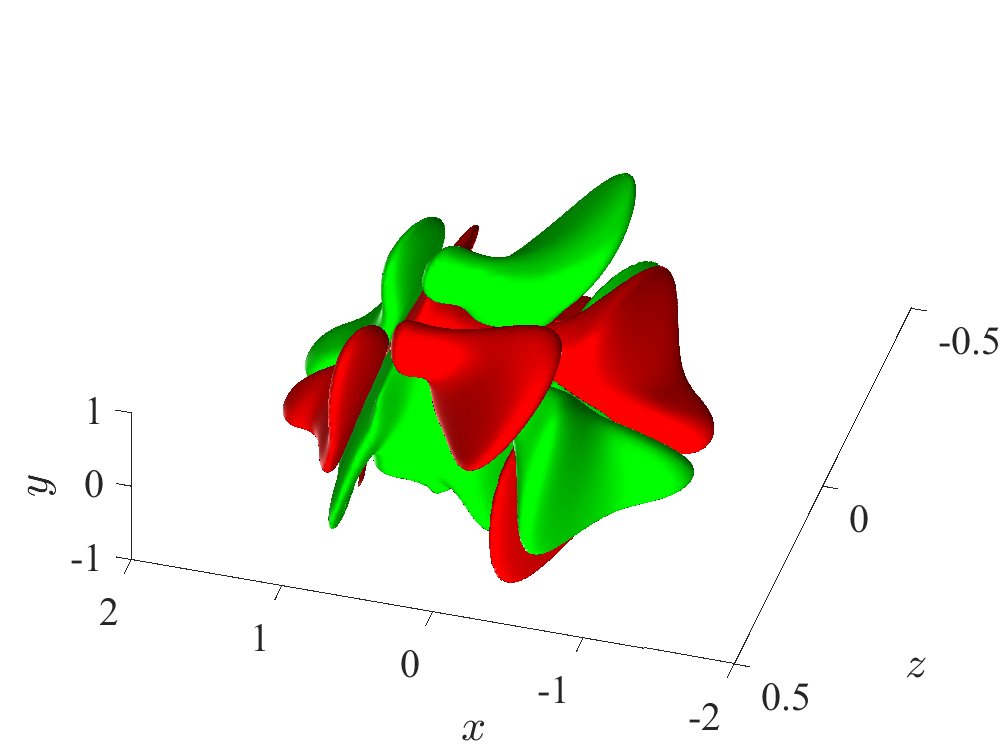}\label{fig:Isoz_c}}\qquad
\subfloat[][$t = 3.5$, Newtonian]{\includegraphics[scale = 0.36]{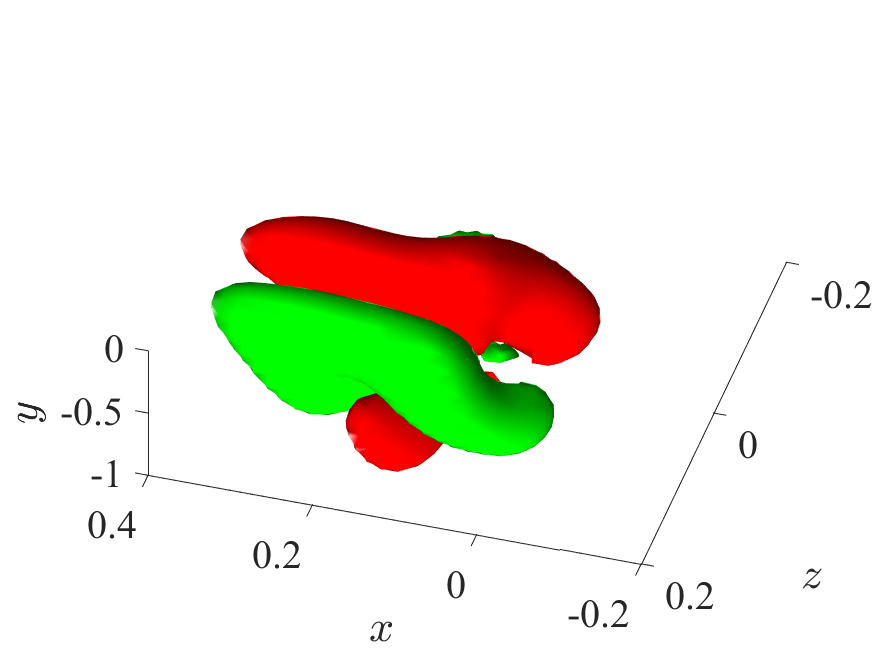}\label{fig:IsozN_c}}\\
\subfloat[][$t = 6$, Viscoelastic]{\includegraphics[scale = 0.36]{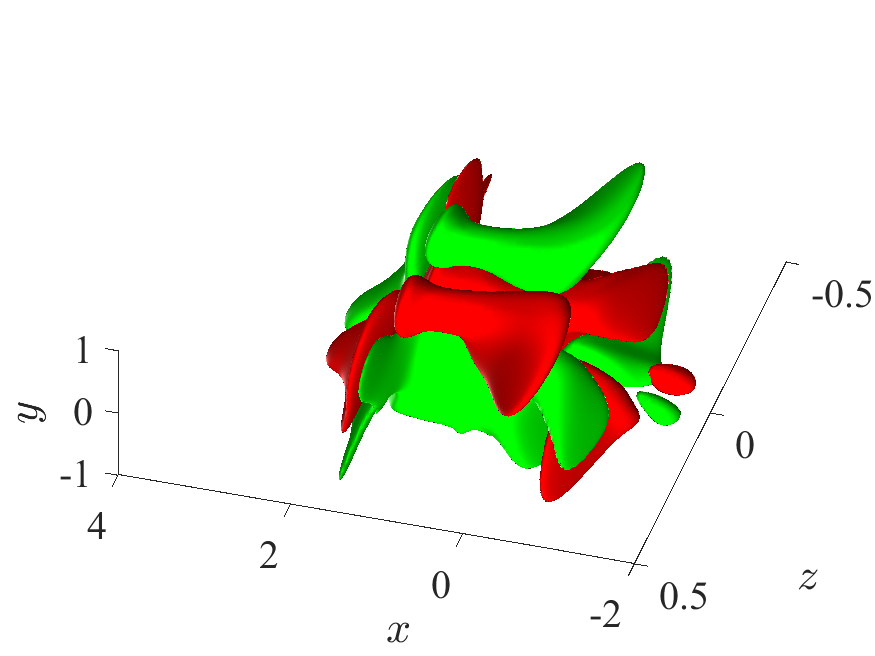}\label{fig:Isoz_d}}\qquad
\subfloat[][$t = 6$, Newtonian]{\includegraphics[scale = 0.36]{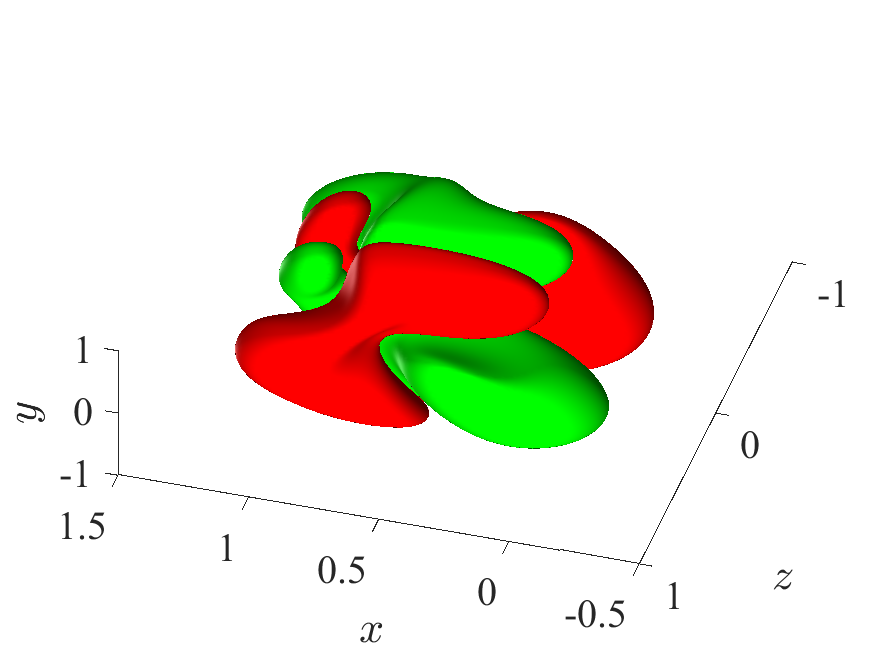}\label{fig:IsozN_d}}
\caption{\label{fig:Isoz} Isosurface plots of the streamwise velocity at $\pm u_{z,\max}/10$ at $Re = 50$. Red color denotes regions of high velocity and green color denotes regions of low velocity. Panels correspond to a viscoelastic fluid at (a) $0.1$, (c) $1$, (e) $3.5$, and (g) $6$ time units, and a Newtonian fluid at (b) $0.1$, (d) $1$, (f) $3.5$, and (h) $6$ time units, with parameters $L = 100$, $\beta = 0.5$ and $\mu = 1$ for the viscoelastic fluid.}
\end{figure}

We now examine flow structures that result from an impulsive excitation in the spanwise direction. Flow structures in physical space provide insight into patterns that result from a localized point force and can suggest potential mechanisms that govern the initial stages of transition to elastic turbulence at low Reynolds numbers. Flow structures presented here are obtained by the pseudospectral method described in \S~\ref{numer_meth}. Time series of flow structures can thus be interpreted as direct numerical simulations of the linearized FENE-CR fluid with an impulsive forcing. As described in \S~\ref{numer_meth}, time stepping procedures are avoided by exploiting linearity to directly obtain flow structures at a given time from the matrix exponential.

Figure \ref{fig:Isoz} shows three-dimensional isosurface plots of the streamwise velocity resulting from an impulsive excitation in the spanwise direction ($u_z$) at the optimal location $y_0 = -0.75$ for the viscoelastic fluid. Figures \ref{fig:Isoz_a}, \ref{fig:Isoz_b}, \ref{fig:Isoz_c} and \ref{fig:Isoz_d} show the time-evolution for a viscoelastic fluid, and Figures \ref{fig:IsozN_a}, \ref{fig:IsozN_b}, \ref{fig:IsozN_c} and \ref{fig:IsozN_d} show the time-evolution for a Newtonian fluid. As discussed in \S~\ref{sec:Energy_Amplification}, in Figure \ref{fig:7c} we observe that viscoelastic fluids produce more oblique structures ($k_x \approx 10^{-1}$, $k_z\approx 10^0$) than Newtonian fluids (in Figure \ref{fig:7f} the streamwise-constant structures with ($k_x \approx 10^{-4}$, $k_z\approx 10^0$) are most amplified). In Figures \ref{fig:Isoz_a} and \ref{fig:Isoz_b}, we see that at early times, the fluctuations in  viscoelastic fluids are more oblique, showing a wavy nature in all three directions. At later times (Figures \ref{fig:Isoz_c} and \ref{fig:Isoz_d}) the wave packet stretches out in the streamwise direction and also spreads across the channel in the wall-normal direction. In contrast, the impulse-induced wave packet in the Newtonian fluid diffuses in space slowly (by observing the scales in the $x$-axis) with a slight amount of translation in the streamwise direction. (Videos of the time-evolution can be found in the supplementary material.)
	
\begin{figure}
\centering
\subfloat[][$t = 0.1$, Viscoelastic (top view)]{\includegraphics[scale = 0.31]{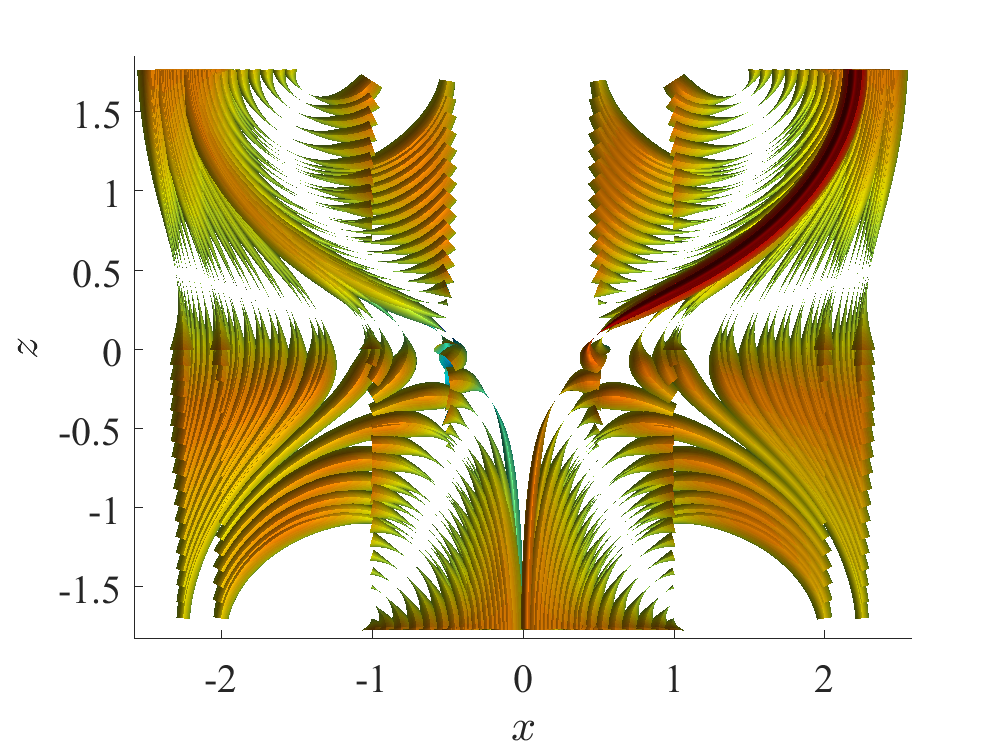}\label{fig:streamz_a1}}\qquad
\subfloat[][$t = 0.1$, Newtonian (top view)]{\includegraphics[scale = 0.31]{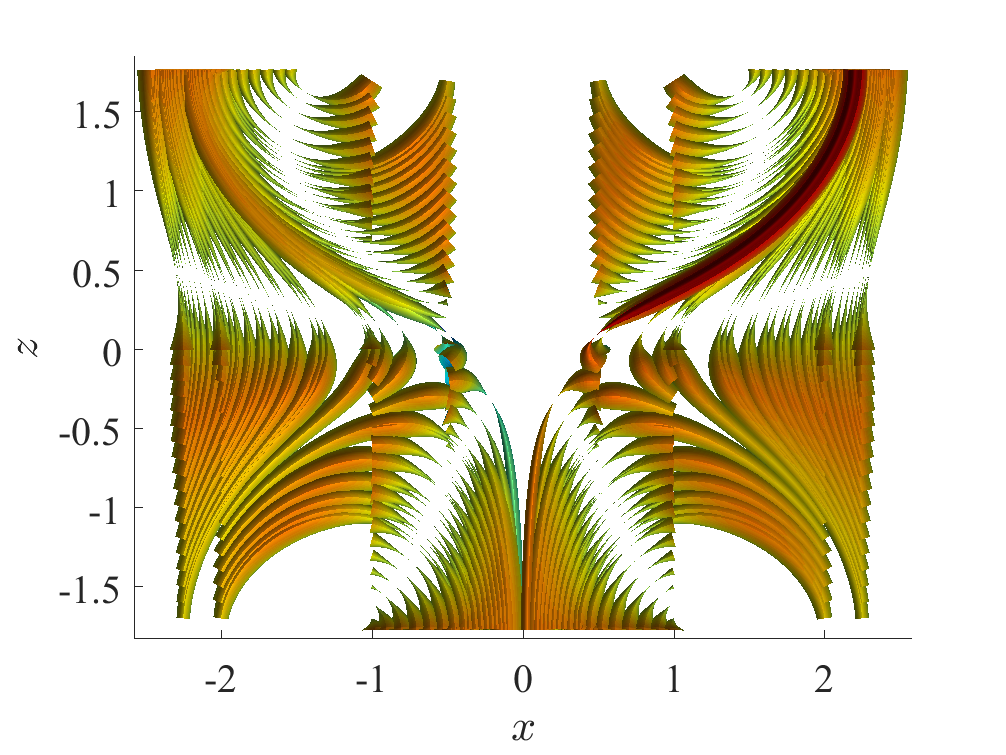}\label{fig:streamzN_a1}}\\
\subfloat[][$t = 0.1$, Viscoelastic]{\includegraphics[scale = 0.33]{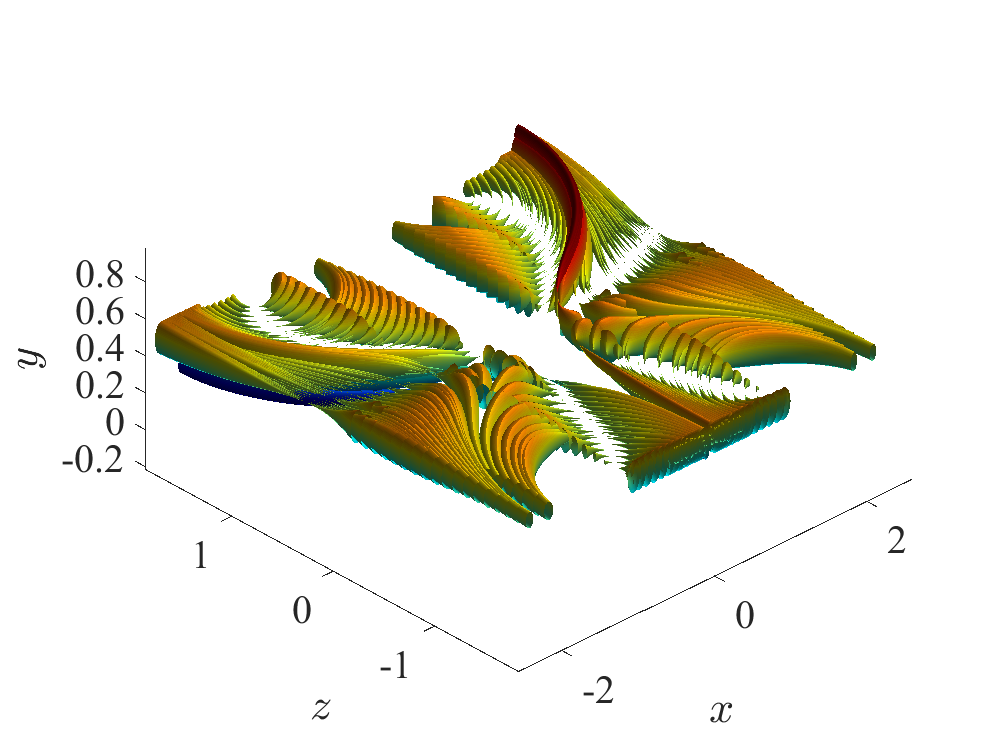}\label{fig:streamz_b1}}\qquad
\subfloat[][$t = 0.1$, Newtonian]{\includegraphics[scale = 0.33]{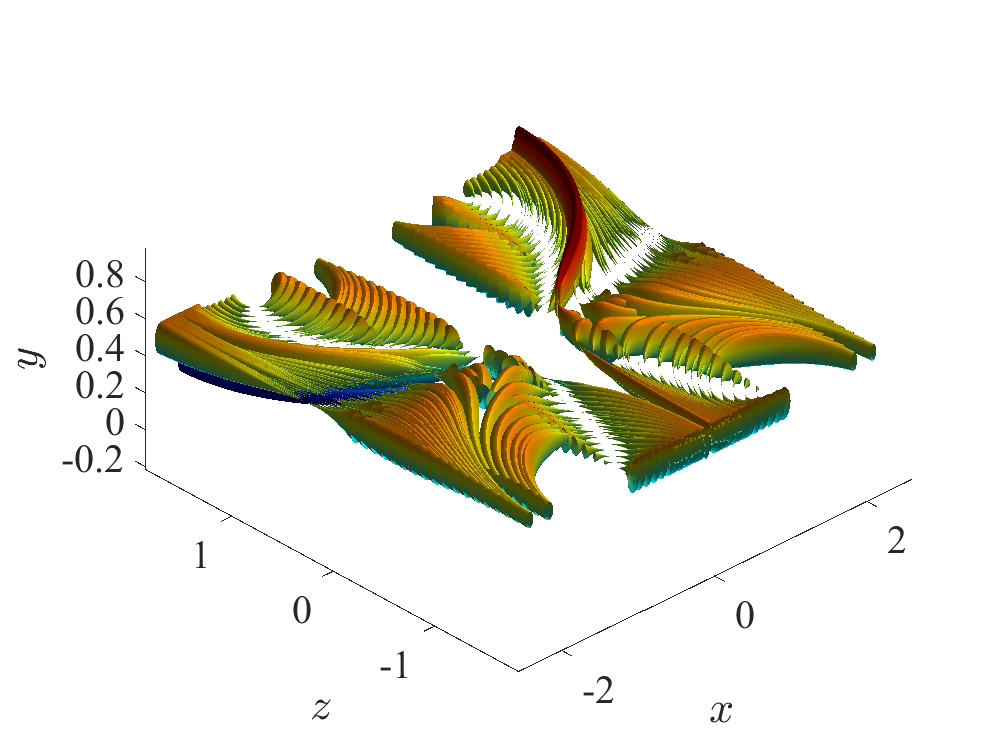}\label{fig:straemzN_b1}}\\

\subfloat[][$t = 6$, Viscoelastic (top view)]{\includegraphics[scale = 0.31]{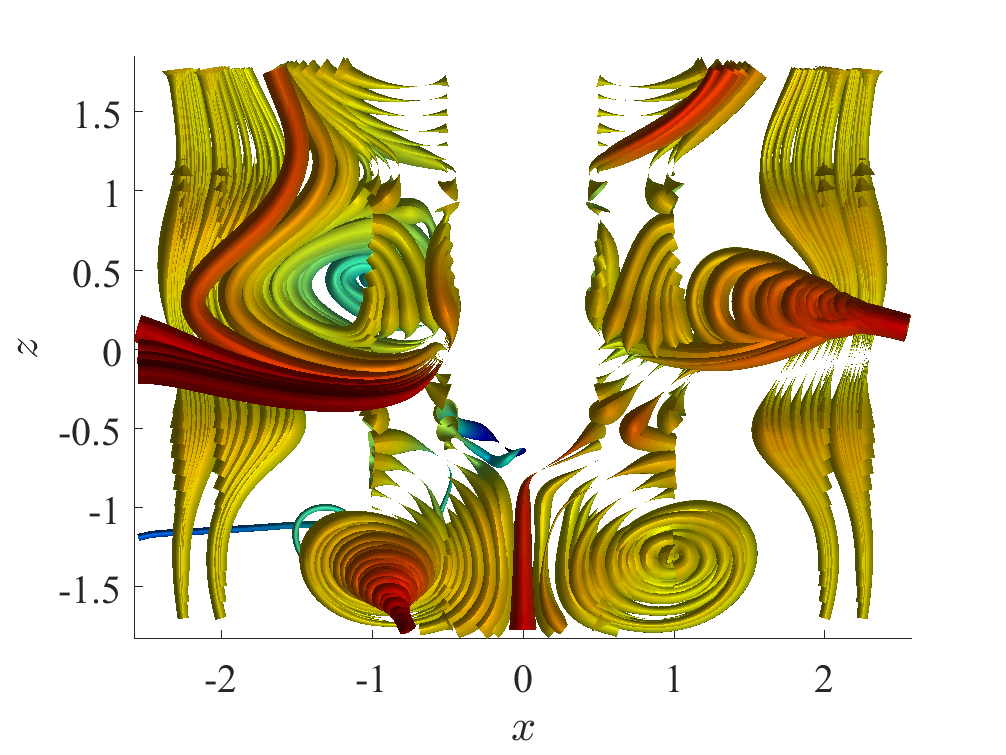}\label{fig:streamz_a}}\qquad
\subfloat[][$t = 6$, Newtonian (top view)]{\includegraphics[scale = 0.31]{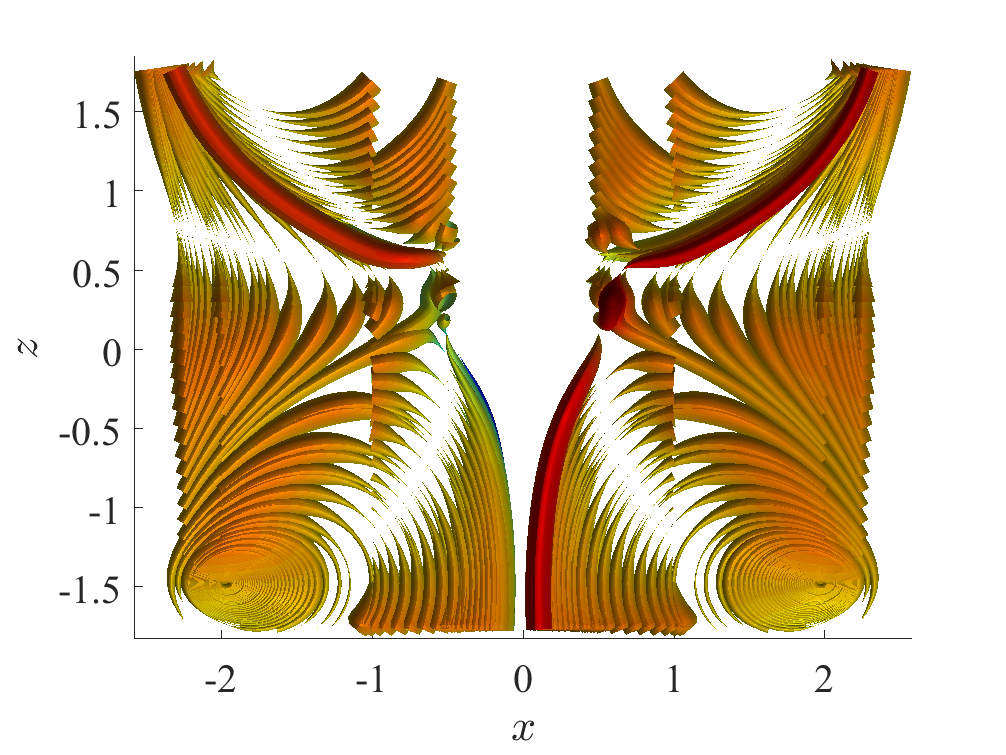}\label{fig:streamzN_a}}\\
\subfloat[][$t = 6$, Viscoelastic]{\includegraphics[scale = 0.33]{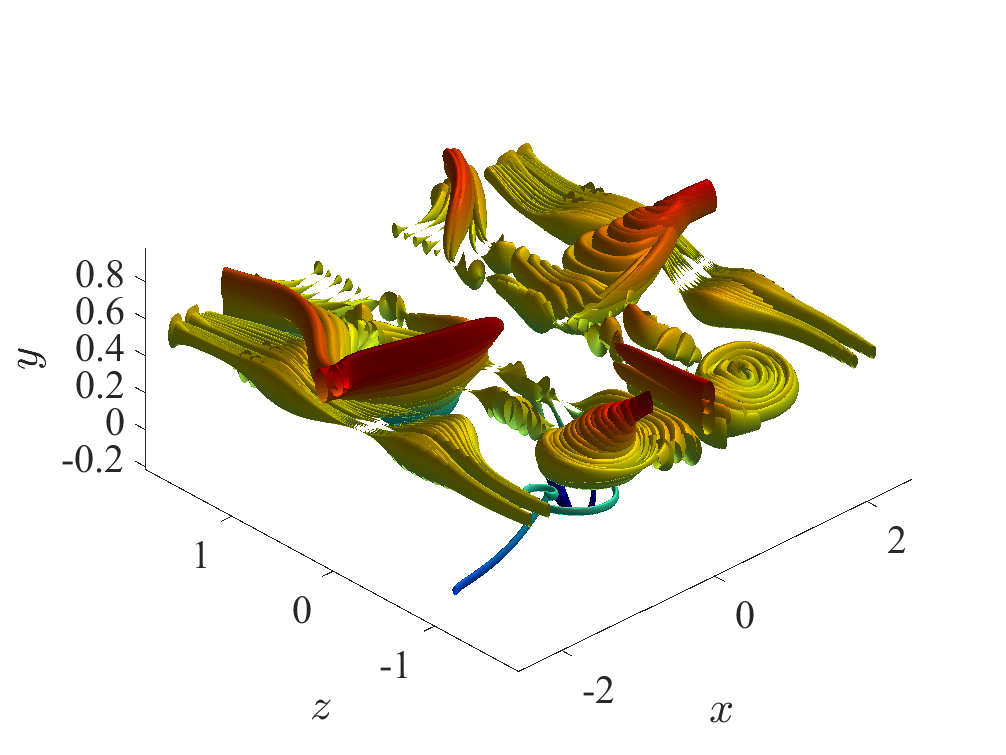}\label{fig:streamz_b}}\qquad
\subfloat[][$t = 6$, Newtonian]{\includegraphics[scale = 0.33]{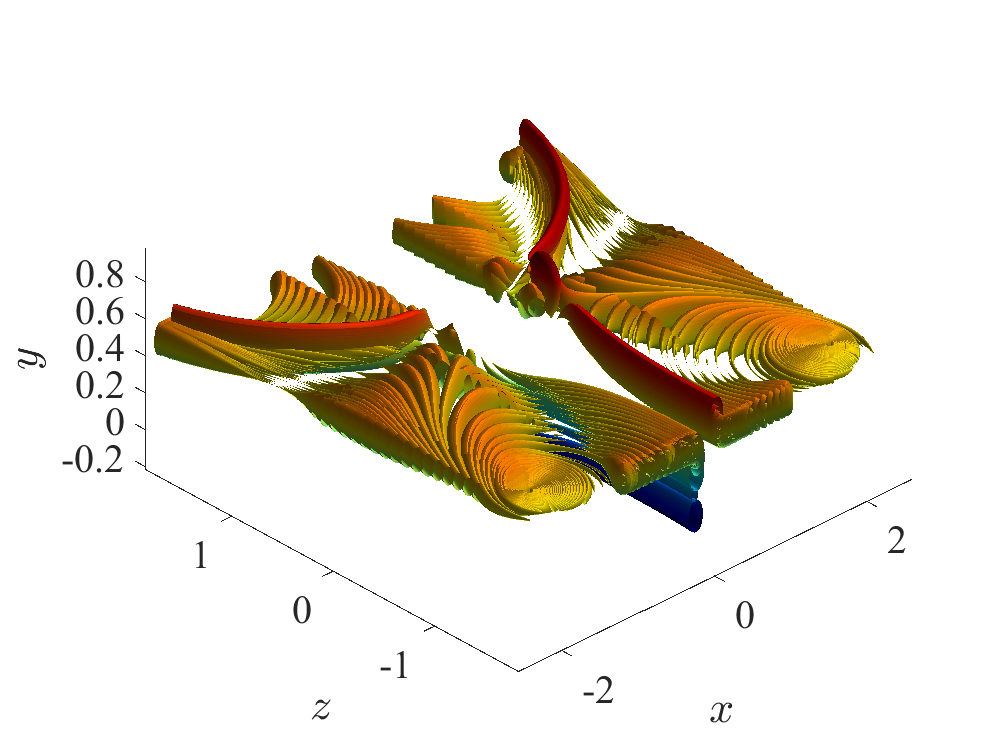}\label{fig:straemzN_b}}\\
\caption{\label{fig:streamz} Three-dimensional streamtubes of the velocity fluctuation vector that originate from the plane $y = 0.5$ at $Re = 50$ with a spanwise impulsive excitation at $y_0 = -0.75$. Show are (a) top view and (c) isometric view for a viscoelastic fluid and (b) top view and (d) isometric view for a Newtonian fluid at $t = 0.1$, and (e) top view and (g) isometric view for a viscoelastic fluid and (f) top view and (h) isometric view for a Newtonian fluid at $t = 6$. Parameters used for the viscoelastic fluid are $L = 100$, $\beta = 0.5$ and $\mu = 1$.}
\end{figure}

Flow structures can be further analyzed by examining three-dimensional streamtubes of the velocity fluctuation vector. Figure \ref{fig:streamz} shows three-dimensional streamtubes that originate from the plane $y=0.5$ for a viscoelastic fluid and a Newtonian fluid. The location $y = 0.5$ is far from the source of the impulse ($y_0 = -0.75$). At the location of the point force, the impulsive excitation is equivalent to an initial condition on the wall-normal velocity and vorticity. This can be seen by considering the general solution of a linear system of the form given in \eqref{eq:12}, for an initial condition $\psi_0(\bkappa)$ with zero forcing ($F_i=0$)~\cite{hespanha2009linear},
\begin{equation}\label{eq:initial_cond}
\phi_i(\bkappa,t) \; = \; {C} \, \mathrm{e}^{At} \,\psi_0(\bkappa).
\end{equation}
The solution of the system with an initial condition in \eqref{eq:initial_cond} and the solution with an impulse forcing given in \eqref{16b} are equivalent if we choose an initial condition $\psi_0$ such that $\psi_0 = F_i(\bkappa)$. We note that $\psi$ is the discrete approximation to $\bpsi = [\BB{r}^T \text{  } v \text{  }\eta]^T$, where $\BB{r}^T$ represents the vector of the six components of the fluctuations of the (symmetric) conformation tensor, $v$ is the wall-normal velocity, and $\eta$ is the wall-normal vorticity. Thus, the impulsive forcing corresponds to an initial condition on the wall-normal velocity and vorticity. This initial condition produces vortical structures even in Newtonian fluids. The interesting feature here is the evolution of vortical structures away from the location of the point force for the viscoelastic fluid. 
 
Figures \ref{fig:streamzN_a1} and \ref{fig:straemzN_b1} show the top and isometric views of streamtubes  for a Newtonian fluid, and Figures \ref{fig:streamz_a1} and \ref{fig:streamz_b1} show the top and isometric views for a viscoelastic fluid at $t = 0.1$. The streamtubes for the Newtonian and  viscoelastic fluids are very similar at $t = 0.1$ as both the Newtonian and viscoelastic fluids have the same initial condition, and at early times the impulse does not significantly contaminate regions away from its source. Figures \ref{fig:streamzN_a} and \ref{fig:straemzN_b} show the top and isometric views of streamtubes  for a Newtonian fluid, and Figures \ref{fig:streamz_a} and \ref{fig:streamz_b} show the top and isometric views for a viscoelastic fluid at $t = 6$. We see in Figure \ref{fig:streamz_a} and \ref{fig:streamz_b} that the viscoelastic fluid generates two pairs of counter-rotating vortices at $x = \pm 1$ that spread out in the wall-normal direction with an oblique inclination. 

In contrast, we do not find significant evolution of vortical structures in the Newtonian fluid  (Figures \ref{fig:streamzN_a} and \ref{fig:straemzN_b}). In fact, streamtubes for the Newtonian fluid at $t = 6$  (Figures \ref{fig:streamzN_a} and \ref{fig:straemzN_b}) are almost the same as they were at $t = 0.1$  (Figure \ref{fig:streamzN_a1} and \ref{fig:straemzN_b1}) (videos of the time-evolution can be found in the supplementary material). The time-evolution of the vortical structures observed here is therefore a unique feature of viscoelastic fluids. Vortex breakdown is a well-known mechanism for transition to turbulence in Newtonian fluids at high Reynolds numbers \cite{hallVortex}. Analyzing the existence of a breakdown and corresponding transition cannot be captured by the linearized dynamics and requires careful consideration of nonlinear effects. However, using the analysis of the linearized dynamics, we find the development of vortical structures that may be related to the initial stages of transition to elastic turbulence. 

Curved streamlines are known to be unstable to small-amplitude perturbations in viscoelastic fluids and growth of these perturbations could eventually lead to elastic turbulence~\cite{larson1990purely,PakdelCurvedstream,MOROZOV2007112}. Here, we find that curved streamlines are generated by an impulsive excitation. If these grow to finite-amplitude, they may also become unstable. Recent work~\cite{pan2013nonlinear} suggests that finite-amplitude perturbations in straight-channel flows can induce elastic turbulence. A potential reason for this transition mechanism may be related to the generation of curved streamlines by nonmodal amplification of  initially small-amplitude disturbances. 

	\vspace*{-3ex}
\section{Concluding remarks}\label{sec:conclusions}
	\vspace*{-1ex}

In this work, we have examined the response of a viscoelastic fluid to localized point forces. We analyzed the kinetic energies of the velocity fluctuations and identified an optimal location to trigger the impulse. The impulse in the optimal location has the maximum impact in the channel and was found to be located near the wall for the viscoelastic fluid. Our analysis has also demonstrated that viscoelastic fluids are more sensitive to small amplitude disturbances when compared to Newtonian fluids at low Reynolds numbers.

Our analysis of kinetic energy showed that the impulse in the spanwise direction has the maximum impact and that the streamwise velocity is most affected. We also found that the amount of amplification increases with increasing elasticity ($\We$) and by increasing polymer concentration ($1-\beta$). These observations agree with earlier studies on the amplification of unstructured channel-wide disturbances~\cite{jovkumPOF10,jovkumJNNFM11,liejovkumJFM13}; the latter observation is also qualitatively consistent with recent experiments~\cite{nolan2016viscoelastic}. The optimal location and direction of the impulse as well as the variation of kinetic energy with polymer concentration and relaxation time studied in this work may provide useful guidelines for inducing elastic turbulence in microfluidic devices and other experiments concerning elastic turbulence. 

We have also shown the spatio-temporal evolution of the wave packet arising from the impulsive excitation. We have demonstrated that the wave packet in the viscoelastic fluid stretches in the streamwise direction. This is in contrast to its Newtonian counterpart which predominantly diffuses in space as a function of time. Three-dimensional streamtubes also revealed time-evolving vortical structures that were not as pronounced in Newtonian fluids. We note that this feature was not observed in previous studies with distributed channel-wide body forces~\cite{hodjovkumJFM08,hodjovkumJFM09,jovkumPOF10,jovkumJNNFM11,liejovkumJFM13}. These structures may provide a mechanism for triggering the initial stages of transition to elastic turbulence in dilute polymer solutions.  

Our results may also be helpful in understanding viscoelastic channel flows that contain a finite-sized object, where the object exerts a drag force on the fluid.  Since any spatially varying and temporally distributed force can be expressed as a summation of impulses, our results may be useful for interpreting the behavior of these more complex flows.  An examination of the nonlinear evolution of fluctuations arising from localized point forces is the next natural step toward addressing the challenging problem of transition to elastic turbulence in viscoelastic channel flows.

	
	\vspace*{-2ex}
\section*{Acknowledgments}

This work is supported in part by the National Science Foundation under grant number CBET-1510654. The Minnesota Supercomputing Institute (MSI) at the University of Minnesota is gratefully acknowledged for providing computing resources.

\appendix
	
	\vspace*{-2ex}
\section{Operators in Poiseuille flow of FENE-CR fluids}\label{appA}
In this section, we define the underlying operators that appear in \eqref{eq:11} in \S~\ref{geef}. We use a Fourier transform defined by
\begin{equation}\label{FourierTrans}
X(\bkappa,y,t) = \int_{-\infty}^\infty\int_{-\infty}^\infty X(x,y,z,t) \, \mathrm{e}^{-j(xk_x+zk_z)}
	\, \mathrm{d}x \, \mathrm{d}z. 
\end{equation}
The operator $\mathbf{A}$ in \eqref{eq:11} is given by
\begin{equation}
\mathbf{A} = 
\left[
\begin{array}{ccc}
\mathcal{R} & \mathcal{V} & \mathcal{N}\\
\mathcal{P}^T & \mathcal{L}_{OS} & 0\\
\mathcal{Q}^T & \mathcal{C}_p & \mathcal{S}
\end{array}
\right].
\end{equation}
Each of the operators that make up $\mathbf{A}$ are given below in Fourier space. The operator $\MM{R}$ is defined as
\begin{equation}
\MM{R} = \left[
\begin{array}{cc}
\MM{R}_{11} & \MM{R}_{12}\\
0_{3\times3} & \MM{R}_{22}\\
\end{array}
\right],
\end{equation}
where
\begin{equation*}
\MM{R}_{11} = \left(
\begin{array}{ccc}
 -\bar{f}/\We-2 \We \left(\bar{U}'\right)^2/\bar{L}^2-j \bar{U} k_x &
   2 \bar{U}' & 0 \\
 -\bar{f} \bar{U}'/\bar{L}^2 & -\bar{f}/\We-j \bar{U} k_x & 0 \\
 0 & 0 & -\bar{f}/\We-j \bar{U} k_x \\
\end{array}
\right),
\end{equation*}

\begin{equation*}
\MM{R}_{12} = \left(
\begin{array}{ccc}
 -2 \We \left(\bar{U}'\right)^2/\bar{L}^2 & 0 & -2 \We
   \left(\bar{U}'\right)^2/\bar{L}^2 \\
 \bar{U}'-\bar{f} \bar{U}'/\bar{L}^2 & 0 & -\bar{f} \bar{U}'/\bar{L}^2 \\
 0 & \bar{U}' & 0 \\
\end{array}
\right),
\end{equation*}
\begin{equation*}
\MM{R}_{22} = \left(
\begin{array}{ccc}
 -\bar{f}/\We-j \bar{U} k_x & 0 & 0 \\
 0 & -\bar{f}/\We-j \bar{U} k_x & 0 \\
 0 & 0 & -\bar{f}/\We-j \bar{U} k_x \\
\end{array}
\right).
\end{equation*}
The operator $\MM{V}$ is defined as 
\begin{equation}\label{apV}
\mathcal{V} = \mathcal{V}_1 + \mathcal{V}_2\;\partial_y +  \mathcal{V}_3\;\partial_{yy}, 
\end{equation}
where $\MM{V}_1$, $\MM{V}_2$, and $\MM{V}_3$ are given by
\begin{equation*}
\MM{V}_1 = 
\left(
\begin{array}{c}
 \left(4 \We^2 \bar{U}' \left(\bar{U}' \bar{f}'-\bar{f} \bar{U}''\right)\right)/\bar{f}^3 \\
 \left(\We \bar{U}' \left(\bar{f}'+2 j \We k_x \bar{U}'\right)+j \bar{f}^2
   k_x-\We \bar{f} \bar{U}''\right)/\bar{f}^2 \\
 0 \\
 \left(2 j \We k_x \bar{U}'\right)/\bar{f} \\
 j k_z \\
 0 \\
\end{array}
\right),
\end{equation*}

\begin{equation*}
\MM{V}_2 = \left(
\begin{array}{c}
 -\left(2 k_x^2 \left(\bar{f}^2+2 \We^2 \left(\bar{U}'\right)^2\right)\right)/(k^2
   \bar{f}^2) \\
 \left(\We \bar{U}' \left(-k_x^2+k^2\right)\right)/(k^2 \bar{f}) \\
- \left(2 k_x k_z \left(\bar{f}^2+\We^2 \left(\bar{U}'\right)^2\right)\right)/(k^2
   \bar{f}^2) \\
 2 \\
 -\left(\We k_x k_z \bar{U}'\right)/(k^2 \bar{f}) \\
- \left(2 k_z^2\right)/(k^2) \\
\end{array}
\right),
\end{equation*}

\begin{equation*}
\MM{V}_3 = \left(
\begin{array}{c}
 \left(2 j \We k_x \bar{U}'\right)/(k^2 \bar{f}) \\
 \left(j k_x\right)/(k^2) \\
 \left(j \We k_z \bar{U}'\right)/(k^2 \bar{f}) \\
 0 \\
 (j k_z)/(k^2) \\
 0 \\
\end{array}
\right),
\end{equation*}
where $k^2 = k_x^2+k_z^2$.
The operator $\MM{N}$ is defined as
\begin{equation}\label{apN}
\mathcal{N} =\mathcal{N}_1+\mathcal{N}_2\;\partial_y,
\end{equation}
where 
\begin{equation*}
\MM{N}_1 = \left(
\begin{array}{c}
 \left(2 k_x k_z \left(\bar{f}^2+2 \We^2 \left(U'\right)^2\right)\right)/(k^2
   \bar{f}^2) \\
 \left(\We k_x k_z \bar{U}'\right)/(k^2 \bar{f}) \\
 -\left(\bar{f}^2 \left(k_x^2-k_z^2\right)+2 \We^2 k_x^2
   \left(U'\right)^2\right)/(k^2 \bar{f}^2) \\
 0 \\
- \left( \We k_x^2 \bar{U}'\right)/(k^2 \bar{f}) \\
- \left(2 k_x k_z\right)(k^2) \\
\end{array}
\right),
\end{equation*}
\begin{equation*}
\MM{N}_2 = \left(
\begin{array}{c}
 -\left(2 j \We k_z \bar{U}'\right)/(k^2 \bar{f}) \\
 -\left(j k_z\right)/(k^2) \\
 \left(j \We k_x \bar{U}'\right)/(k^2 \bar{f}) \\
 0 \\
 (j k_x)/(k^2) \\
 0 \\
\end{array}
\right).
\end{equation*} 
The operator $\MM{P}$ is defined as
\begin{equation}\label{apP}
\mathcal{P}^T = \frac{(1-\beta)}{Re}\Delta^{-1}\left(\mathcal{P}_1^T + \mathcal{P}_2^T\;\partial_y + \mathcal{P}_3^T\;\partial_{yy}\right),
\end{equation}
where $\Delta = \partial_{yy}-k^2$ is the Laplacian in Fourier space, and
\begin{equation*}
\MM{P}_1 = \left(
\begin{array}{c}
 \left(k_x \left(\bar{f}' \left(\bar{L}^2 k_x-2 j \We \bar{U}''\right)+\We \bar{U}'
   \left(4 \We k_x \bar{U}''-j \left(k^2
   \bar{f}+\bar{f}''\right)\right)\right)\right)/(\We \bar{L}^2) \\
 -\left(j k_x \left(k^2 \bar{f}+\bar{f}''\right)\right)/\We \\
 \left(2 k_x k_z \bar{f}'\right)/\We\\
 \left(-\bar{f}' \left(k^2 \bar{L}^2+2 j \We k_x \bar{U}''\right)-\We k_x \bar{U}'
   \left(j k^2 \bar{f}+j \bar{f}''-4 \We k_x \bar{U}''\right)\right)/(\We \bar{L}^2)
   \\
 -\left(j k_z \left(k^2 \bar{f}+\bar{f}''\right)\right)/(\We) \\
 \left(\bar{f}' \left(\bar{L}^2 k_z^2-2 j \We k_x \bar{U}''\right)-\We k_x \bar{U}'
   \left(j k^2 \bar{f}+j \bar{f}''-4 \We k_x \bar{U}''\right)\right)/(\We \bar{L}^2)
   \\
\end{array}
\right),
\end{equation*}
\begin{equation*}
\MM{P}_2 = \left(
\begin{array}{c}
 \left(k_x \left(\bar{f} \left(\bar{L}^2 k_x-2 j \We \bar{U}''\right)+2 \We \bar{U}'
   \left(\We k_x \bar{U}'-j \bar{f}'\right)\right)\right)/(\We \bar{L}^2) \\
 -\left(2 j k_x \bar{f}'\right)/\We \\
 \left(2 \bar{f} k_x k_z\right)/\We \\
 \left(2 \We k_x \bar{U}' \left(\We k_x \bar{U}'-j \bar{f}'\right)-\bar{f} \left(k^2
   \bar{L}^2+2 j \We k_x \bar{U}''\right)\right)/(\We \bar{L}^2) \\
 -\left(2 j k_z \bar{f}'\right)/\We \\
 \left(\bar{f} \left(\bar{L}^2 k_z^2-2 j \We k_x \bar{U}''\right)+2 \We k_x \bar{U}'
   \left(\We k_x \bar{U}'-j \bar{f}'\right)\right)/(\We \bar{L}^2) \\
\end{array}
\right),
\end{equation*}
\begin{equation*}
\MM{P}_3 = \left(
\begin{array}{c}
 -\left(j \bar{f} k_x \bar{U}'\right)/(\bar{L}^2) \\
 -(j \bar{f} k_x)/\We \\
 0 \\
 -(j \bar{f} k_x \bar{U}')/\bar{L}^2 \\
 -(j \bar{f} k_z)/\We \\
 -(j \bar{f} k_x \bar{U}')/\bar{L}^2 \\
\end{array}
\right).
\end{equation*}
The operator $\MM{Q}$ is defined as
\begin{equation}\label{eq:apQ}
\mathcal{Q} = \frac{(1-\beta)}{Re}\left(\mathcal{Q}_1 + \mathcal{Q}_2\;\partial_y\right),
\end{equation}
where
\begin{equation*}
\MM{Q}_1 = \left(
\begin{array}{c}
 \left(k_z \left(\bar{f} \left(j \We \bar{U}''-\bar{L}^2 k_x\right)+\We \bar{U}'
   \left(j \bar{f}'-2 \We k_x \bar{U}'\right)\right)\right)/(\We \bar{L}^2) \\
 \left(j k_z \bar{f}'\right)/\We \\
 \left(\bar{f} \left(k_x-k_z\right) \left(k_x+k_z\right)\right)/\We \\
 \left(k_z \left(j \bar{f} \bar{U}''+j \bar{U}' \bar{f}'-2 \We k_x
   \left(\bar{U}'\right)^2\right)\right)/\bar{L}^2 \\
 -\left(j k_x \bar{f}'\right)/\We \\
 \left(k_z \left(\bar{f} \left(\bar{L}^2 k_x+j \We \bar{U}''\right)+\We \bar{U}'
   \left(j \bar{f}'-2 \We k_x \bar{U}'\right)\right)\right)/(\We \bar{L}^2) \\
\end{array}
\right),
\end{equation*}
\begin{equation*}
\MM{Q}_2 = \left(
\begin{array}{c}
 \left(j \bar{f} k_z \bar{U}'\right)/\bar{L}^2 \\
 \left(j \bar{f} k_z\right)/\We \\
 0 \\
 (j \bar{f} k_z \bar{U}')/\bar{L}^2\\
 -(j \bar{f} k_x)/\We \\
 (j \bar{f} k_z \bar{U}')/\bar{L}^2 \\
\end{array}
\right).
\end{equation*}
The operators $\MM{L}_{OS}$, $\MM{C}_p$ and $\MM{S}$ are the Orr-Sommerfeld, coupling, and Squire operators respectively,
\begin{equation}\label{apLosCpS}
\begin{split}
\MM{L}_{OS} &= \Delta^{-1}\left(-j k_x \bar{U} \Delta \, + \, j k_x \bar{U}'' \, + \, \frac{\beta}{Re}\, \Delta^2\right),\\
\MM{C}_p &= -j k_z \bar{U}',\\
\MM{S} &= -jk_x \bar{U} \, + \, \frac{\beta}{Re} \, \Delta.
\end{split}
\end{equation}
The operator $\mathbf{B}$ in \eqref{eq:11} is given by
\begin{equation}\label{eq:apB}
\mathbf{B} = \left[
\begin{array}{c}
0_{6\times3}\\
\mathbf{B}_1\\
\mathbf{B}_2\\
\end{array}
\right],
\end{equation}
where $0_{6\times 3}$ represents a $6\times 3$ matrix of zeros, and
\begin{equation*}
\mathbf{B}_1 = \Delta^{-1}\left[
\begin{array}{c}
-jk_x\, \partial_y\\
-k^2 I \\
-jk_z\, \partial_y\\
\end{array}
\right]^T,\quad
\mathbf{B}_2 = \left[
\begin{array}{c}
jk_z I \\
0\\
-jk_x I \\
\end{array}
\right]^T.
\end{equation*}
The operator $\mathbf{C}$ in \eqref{eq:11} is given by
\begin{equation}\label{apC}
\mathbf{C} 
	\; = \, 
\left[
\begin{array}{cc}
\mathbf{C}_u\\
\mathbf{C}_v\\
 \mathbf{C}_w\\
\end{array}
\right]
	\, = \,
\left[
\begin{array}{c}
\mathbf{C}_u\\
\mathbf{C}_v\\
\mathbf{C}_w\\
\end{array}
\right] 
	\, = \; \dfrac{1}{k^2}\left[
\begin{array}{ccc}
0_{1\times6} & jk_x\,\partial_y & -jk_z I \\
0_{1\times6} & k^2 I & 0\\
0_{1\times6} & jk_z\,\partial_y & jk_x I \\
\end{array}
\right].
\end{equation}
where $0_{1\times 6}$ represents a $1\times 6$ submatrix of zeros.

\section{Inner product that determines the kinetic energy}\label{appB}

In this section, we define the inner product that determines the kinetic energy of fluctuations  discussed in \S~\ref{KE}.

The Hilbert space for the operator $\mathbf{A}$ (see \eqref{eq:11} in \S~\ref{geef}) can be defined on the basis of its domain and boundary conditions \cite{jovbamJFM05,butler1992three}. We define the space of functions $\mathbb{H}_{OS}$
\begin{equation}
\mathbb{H}_{OS} \, \DefinedAs \, \left\{g \in L^2[-1,1]; \; g'' \in L^2[-1,1]; \; g(\pm1) \, = \,0\right\}.
\end{equation}
The domain of the operators $\mathcal{P}$, $\mathcal{Q}$, $\mathcal{V}$, and $\mathcal{N}$ is $\mathbb{H}_{OS}^{\,6\times 1}$, and the domain of $\MM{S}$ is $\mathbb{H}_{OS}$. The domain of the $\mathcal{L}_{OS}$ can be defined as
\begin{equation}
\mathcal{D}(\mathcal{L}_{OS}) \, \DefinedAs \, \left\{g\in\mathbb{H}_{OS}; \; g''''\in L^2[-1,1]; \; g'(\pm1) \, = \, 0\right\}.
\end{equation}
We define the following weighted inner product for functions $\bxi_1,\bxi_2\in \mathbb{H}_{OS}^{8\times 1}$,
\begin{equation}\label{WeightedinnerProd}
\left<\bxi_1,Q\bxi_2\right> \, \DefinedAs \, \left<\bxi_1,\bxi_2\right>_\mathrm e,
\end{equation}
where $\left<\cdot,\cdot\right>$ is the standard $L^2[-1,-1]$ inner product and $Q$ is a linear operator given by, 
\begin{equation*}
Q \; = \; \lim_{\iota \, \to \, 0^+} \frac{1}{k^2}\left[
\begin{array}{ccc}
\iota I_{6\times6} & 0 & 0\\
0 & -\Delta & 0\\ 
0 & 0 & I\\
\end{array}
\right],
\end{equation*}
where, $I_{m\times n}$ is a block matrix identity operator of dimensions $m$ by $n$. The inner product defined in \eqref{WeightedinnerProd} determines the energy of velocity fluctuations. It can be verified that (see \cite{butler1992three}) that the kinetic energy can be evaluated as 
	\begin{equation}
	\label{ke}
	\left<\bpsi (\bkappa,t),\bpsi (\bkappa,t)\right>_\mathrm e
	\; = \; 
	\left<\bphi_i (\bkappa,t),\bphi_i (\bkappa,t) \right> 
	\; = \; 	
	\int_{-1}^1
	\BB{v}_i^* (\bkappa,y,t) \, \BB{v}_i (\bkappa,y,t)  
	\, 
	\mathrm{d}y,
\end{equation}
where $\bphi$ is the vector of outputs (i.e., the velocity fluctuations $\bphi = [\,u\, ~\, v\, ~\, w\,]^T$) and $\bpsi = [\,\BB{r}^T\, ~\, v\, ~\, \,\eta\,]^T$ is the vector of state variables that appear in state-space representation~\eqref{eq:11} of the FENE-CR model presented in \S~\ref{geef}. 
The adjoints of $\mathbf{A}$,  $\BB{F}_i$, and $\mathbf{C}$ are defined with respect to the inner-product defined in \eqref{WeightedinnerProd} as
\begin{equation}\label{adjoint}
\begin{split}
\left<\psi,\mathbf{A}\psi\right>_\mathrm e &= \left<\mathbf{A}^\dagger\psi,\psi\right>_\mathrm e\\
\left<\psi,\BB{F}_i\,g(t)\right>_\mathrm e &= \left<\BB{F}_i^\dagger\psi,g(t)\right>_\mathbb{C}\\
\left<\phi,\mathbf{C}\psi\right> &= \left<\mathbf{C}^\dagger\phi,\psi\right>_\mathrm e\\
\end{split}
\end{equation}
Here, $\left<\cdot,\cdot\right>_\mathrm e$ is the weighted inner product defined in \eqref{WeightedinnerProd}, $\left<\cdot,\cdot\right>$ is the standard $L^2[-1,-1]$ inner product, and $\left<\cdot,\cdot\right>_\mathbb{C}$ is the standard vector inner product that induces a Euclidean norm.

\end{document}